\documentclass[twocolumn,trackchanges]{aastex61}

\newcommand{\msun}{\rm M_{\sun}}

\usepackage{hyperref}
\usepackage[hyphenbreaks]{breakurl}
\usepackage{graphicx}
\usepackage{subfigure}
\usepackage{tabularx} 
\usepackage{color}
\usepackage{CJK}
\received{XXX}
\revised{\today}
\accepted{XXX}
\submitjournal{ApJ}

\begin{document}
\begin{CJK*}{UTF8}{gbsn}
\title{A Large Massive Quiescent Galaxy Sample at $z\sim1.2$}

\correspondingauthor{Y.Sophia Dai}
\email{ydai@nao.cas.cn}

\author{Hai Xu}
\affil{Chinese Academy of Sciences South America Center for Astronomy (CASSACA), National Astronomical Observatories of China, Chinese Academy of Sciences, 20A Datun Road, Beijing}
\affil{School of Astronomy and Space Sciences, University of Chinese Academy of Sciences, Beijing 100049, China}
\author{Y.Sophia Dai(戴昱)}
\author{Jia-Sheng Huang}
\affil{Chinese Academy of Sciences South America Center for Astronomy (CASSACA), National Astronomical Observatories of China, Chinese Academy of Sciences, 20A Datun Road, Beijing}

\author{Zhaoyu Wang}
\affil{Department of Astronomy, School of
 Physics and Astronomy, and Shanghai Key Laboratory for
Particle Physics and Cosmology, Shanghai Jiao Tong University,
Shanghai 200240, China}

\author{Cheng Cheng}
\author{Xu Shao}
\author{Shumei Wu}
\affil{Chinese Academy of Sciences South America Center for Astronomy (CASSACA), National Astronomical Observatories of China, Chinese Academy of Sciences, 20A Datun Road, Beijing}

\author{Xiaohu Yang}
\author{Yipeng Jing}
\affil{Department of Astronomy, School of
 Physics and Astronomy, and Shanghai Key Laboratory for
Particle Physics and Cosmology, Shanghai Jiao Tong University,
Shanghai 200240, China}

\author[0000-0002-7712-7857]{Marcin Sawicki}
\altaffiliation{Canada Research Chair}
\affil{Institute for Computational Astrophysics and Department of Astronomy \& Physics, Saint Mary's University, Halifax, N.S.\ B3H 3C3, Canada}
\affil{NRC Herzberg Astronomy and Astrophysics, 5071 West Saanich Road, Victoria, B.C.\ V9E 2E7, Canada}

\author{Feng-Yuan Liu}
\affil{Chinese Academy of Sciences South America Center for Astronomy (CASSACA), National Astronomical Observatories of China, Chinese Academy of Sciences, 20A Datun Road, Beijing}
\affil{School of Astronomy and Space Sciences, University of Chinese Academy of Sciences, Beijing 100049, China}

\begin{abstract}

In this paper we present a simple color-magnitude selection
and obtain a large sample of 33,893 massive quiescent galaxies at intermediate redshifts (1 $<z < $1.5). 
We choose the longest wavelength available in the Hyper-Supreme-Cam (HSC) deep survey, 
the $Y$ band and $i-Y$ color, 
to select the 4000\AA\ Balmer jump in passive galaxies to the highest redshift possible within the survey. 
With the rich multi-wavelength data in the HSC deep fields, 
we then confirm that the selected galaxies are in the targeted redshift range of $1<z<1.5$, 
lie in the passive region of the UVJ diagram,
and have high stellar masses at log(M$_*/\msun)>10.5$, with a median of log(M$_*/\msun)=11.0$.
A small fraction of our galaxies is also covered by the HST CANDELS. 
Morphological analysis in the observed H band 
shows that the majority of this subsample are early-type galaxies.
As massive early-type galaxies trace the high density regions in the large scale structure in the universe,
our study provides a quick and simple way to obtain a statistical significant sample of massive galaxies 
in a relative narrow redshift range. 
Our sample is 7--20 $\times$ larger at the massive end (log(M$_*/\msun)>10.5$)
than any existing samples obtained in previous surveys. 
This is a pioneer study, and the technique introduced here can be applied to future wide-field survey to 
study large scale structure, and to identify high density region and clusters.

\end{abstract}


\section{Introduction} 
Understanding the formation and evolution of massive galaxies (M $\sim$ 10$^{11}\msun$) is 
one of the key questions in extragalactic studies.
The downsizing scenario shows that compared to lower mass galaxies, 
massive galaxies had their star formation 
and evolution happened earlier and assembled their stellar mass more quickly
 \citep{1996AJ....112..839C, 2000ApJ...536L..77B, 2005ApJ...619L.135J, 2007ApJ...665..265F}.
In the cosmic history, $z\sim2$ is an important epoch when
powerful star-forming galaxies dominated the star formation activity, 
assembled most of their stellar mass, 
and formed large scale structures \citep[e.g.][]{2007A&A...468...33E, 2007ApJ...660L..43N,2014ARA&A..52..415M}.
Those galaxies eventually became massive early-type galaxies (ETGs)
in the local universe, yet most of their stellar masses were already in place as early as 
at $z\sim1$ \citep[e.g.][]{2004A&A...424...23F, 2005ApJ...634..861Y, 2006ApJ...651..120B, 
2006A&A...453L..29C, 2006A&A...453..869B, 2007ApJS..172..494S, 2007ApJ...654..858B}.
Quenching of intensive star formation must occur in these powerful star forming galaxies in $1< z <2$
to convert them to passive galaxies  \citep[e.g.][]{2004A&A...424...23F, 2004Natur.430..181G, 2007ApJ...669..184A, 2007A&A...476..137A}.

Many surveys and observations have detected such strong evolution in massive galaxies in 1$< z <2$. 
In the Galaxy Mass Assembly ultra-deep Spectroscopic Survey (GMASS), 
\citet{2008A&A...482...21C} stacked the spectra of 
13 passive galaxies in $1<z<2$ to a total integration time of $\sim$500 hours. 
The stacked spectrum  indicated that
these passive galaxies were very likely in the subsequent stage of powerful starburst galaxies at $z>2$.  
Such an evolution also requires a morphological transition 
as starburst and passive galaxies have different morphologies. 
\citet{2009ApJ...706.1364F} found that star-forming galaxies at $z\sim2$ 
has rotating disks with high velocity dispersion, 
which makes these galaxies unstable and leads to rapid transformation to the spheroids.
The evolution from star-forming galaxies at $z\sim2$ to typical red ellipticals at $z\sim1.2$ 
corresponds to a timescale of $\sim$2Gyr, 
according to the AGN-based quenching model 
\citep[e.g.][]{2006ApJ...647..753M, 2006MNRAS.370..645B, 2008ApJS..175..390H, 2010ApJ...721..193P}. 
There were many studies of passive galaxies at $z>1.2$, 
yet a larger sample with better statistics is needed 
to study their hosting halo properties and their direct link to the star forming galaxies at $z\sim2$. 
 
In the past two decades, 
new generations of wide-field optical$/$infrared cameras,
as well as various photometric selection techniques, 
yielded many massive passive galaxy samples at $z>1$ 
\citep[e.g.][]{2005ApJ...624L..81L, 2005ApJ...635..832M, 2006ApJ...649L..71K, 2007ApJ...654L.107B, 2008ApJ...676..781W, 2009A&A...500..705M, 2009A&A...501...15F, 2010ApJ...725.1277M, 2014ApJ...787L..36S, 2014ApJ...783L..14S, 2017Natur.544...71G, 2018A&A...618A..85S}. 
At higher redshifts up to z$\sim$2, substantial number of massive galaxies have 
already quenched their star formation and became passive \citep[e.g.][]{2013A&A...556A..55I}.
The fraction of passive galaxies at the massive end of M$>10^{10.5}\msun$ increases rapidly, 
from $\sim$ 30\% at $z\sim2$ to $\sim$ 60\% at $z\sim0.5$ 
\citep{2013ApJ...777...18M, 2014ApJ...783...85T, 2017ApJ...847..134K}. 
In the local universe, early-type galaxies (ETGs)
become overwhelmingly dominant at M$>10^{10.5}\msun$ \citep[e.g.][]{2003A&A...402..837P}.
There is also a size evolution for passive galaxies.
Massive passive galaxies at z$>$1 appeared to be more compact
than their local counterparts \citep[e.g.][]{2005ApJ...626..680D, 2008ApJ...677L...5V, 2008ApJ...687L..61B, 2009ApJ...695..101D, 2010ApJ...713..738W}.
The size growth depends on various parameters like mass and morphology,   
though the mass-size relation remains unchanged for galaxies up to $z\sim$2,
with no clear dependence on their environment,
whether they are in clusters or in the field \citep[e.g.][]{2013MNRAS.428.1715H}.

Massive galaxies are also good tracers of high density regions in the universe. 
This is particularly true for ETGs based on the morphology-density relation \citep[e.g.][]{2002MNRAS.332..827N, 2009MNRAS.399..878R}.
In practice, one way of connecting different galaxy populations at different redshifts is 
to compare their bias calculated from the two-point correlation function
\citep[e.g.][]{2007ApJ...671..278G,2013MNRAS.431.3045H,2017MNRAS.464.1380W}.
Some recent surveys found that passive galaxies at $z\sim1$ are strongly clustered 
with a bias ($b_{g}$) $>2$ at a mean halo mass of $M_{\rm halo})\sim$10$^{13}\msun$
 \citep{2013MNRAS.431.3045H, 2017MNRAS.464.1380W}.  
 K band-selected passive galaxies at $z\sim1.6$ were found to have a mean halo mass of $M_{\rm halo}\sim$10$^{13}\msun$ 
 and a bias of $b_{g}=3.27\pm\,0.46$ \citep{2008ApJ...681.1099B}. 
Next generation surveys will probe deeper into the universe 
with much larger coverages, which will yield larger samples and better bias estimates.

There are several ways to select passive galaxies at high redshifts \citep[e.g.][]{2006ARA&A..44..141R}.
Early studies exploited optical-NIR colors \citep[e.g.][]{1988ApJ...331L..77E, 1990ApJ...360L...1C, 1992ApJ...386...52M, 1999MNRAS.308.1061S} to identify the Balmer break in passive galaxies at z$>$1. 
One possible degeneracy in such one-color selection is that 
both dusty and old galaxies can produce red optical-NIR colors \citep[e.g.][]{2002A&A...381L..68C, 2007MNRAS.376.1054D, 2014ApJ...787L..36S, 2016ApJ...827L..25M}.
Multi-wavelength data for galaxies were applied to disentangle this degeneracy.
Furthermore, passive galaxies at z$>$1 are extremely difficult to identify spectroscopically 
due to the lack of emission lines in their spectra. 
Photometric redshifts are essential to study this type of galaxies. 
With known photometric redshifts, one effective method to separate dusty and passive galaxies 
is by their rest-frame $UVJ$ diagram \citep{2005ApJ...624L..81L, 2007ApJ...655...51W, 2009ApJ...691.1879W}. 
At z$>$1, the rest-frame J band is redshifted to the observed Mid-InfraRed(MIR) bands. 
Therefore at z$>$1, multi-wavelength data up to MIR bands are needed to identify passive galaxies
using the UVJ diagram.

It is still very challenging to obtain a large photometry sample for passive galaxies at high redshifts. 
Both deep wide-field optical and IR surveys are needed for photometric redshift measurement and SED classification. 
The largest samples to date are from the COSMOS survey \citep{2013A&A...556A..55I, 2013ApJ...777...18M}.
At 1 $< z < $1.5,  the COSMOS passive galaxy samples
include 6,455 \citep{2013ApJ...777...18M} and 3,796 \citep{2013A&A...556A..55I} passive galaxies
with stellar mass greater than $\sim$ 10$^{9.6}\msun$, 
respectively.
On the other hand, 
spectroscopical studies of high redshift massive passive galaxies
have been done for individual sources \citep{2008A&A...482...21C, 2017Natur.544...71G, 2017ApJ...834...18B, 2019ApJ...874...17B}.  
Unfortunately, due to the long exposures required, 
e.g. several hours are needed for a single detection \citep[e.g.][]{2005ApJ...634..861Y, 2017Natur.544...71G},
it is unlikely to assemble a large spectroscopically identified passive galaxy sample. 
Color-magnitude selection is still the most efficient way 
to construct large samples of massive passive galaxies. 

In this paper, 
we propose a simple color-magnitude selection to select massive passive galaxies
at $z\sim1.2$ in the HSC udeep and deep fields. 
Though multi-wavelength photometry and photometric redshifts
are more effective in the selection of galaxies, 
we will argue in this paper that a single color selection remains a valid method in 
selecting massive passive galaxies in wide field optical surveys.   
We further argue that this single color selection is more straightforward in studying large 
scale structure and mean halo properties, as demonstrated in this and following papers. 
Several wide field optical surveys are planned in the near future, 
and will yield much larger samples, to which we can apply this techniques to study the mass distribution and clustering. 
In this paper, we present a single color selection of passive galaxies at z$\sim$1.2,
from the recent deep survey in the Hyper Suprime-Cam Subaru Strategic Program \citep[HSC-SSP;][]{2018PASJ...70S...8A,2019PASJ...71..114A}. 
This selection yields a sample of more than 33,000 galaxies with stellar mass in 10.5$<log(M_*/\msun)<$11.5, 
almost a factor of 7-15 times larger than any existing passive galaxies samples in the same mass and redshift range. 
These massive galaxies trace the high density regions in the universe, 
which can be used to identify clusters with the same color criterion \citep{2005APJS.157.1G, 2007ASP.634..861Y, 2009ApJ.698.1943W}.

In \S\ref{sec:yband}, we present the sample selection. 
Then we perform the multi-wavelength analysis and morphology classification in \S\ref{sec:validation},
and obtain the redshift distribution and mass function for the sample to further validate our selection.
In \S\ref{sec:hod}, we derive the angular correlation function and halo occupation distribution (HOD).
Our summary is in \S\ref{sec:summary}.
We adopt the AB magnitude system in this paper. 
The adopted initial mass function (IMF) is from \citet{2003PASP..115..763C}.
The paper uses cosmological parameters $h\equiv H_{\rm 0}$[km s$^{-1}$ Mpc$^{-1}$]$/100=0.70$, $\Omega_\Lambda =0.7$, $\Omega_{\rm M} =0.3$.

\section{A Passive Galaxy Sample at z$\sim$1.2} \label{sec:yband}

\subsection{Optical and Infrared Data}

\begin{figure}
\includegraphics[width=0.48\textwidth]{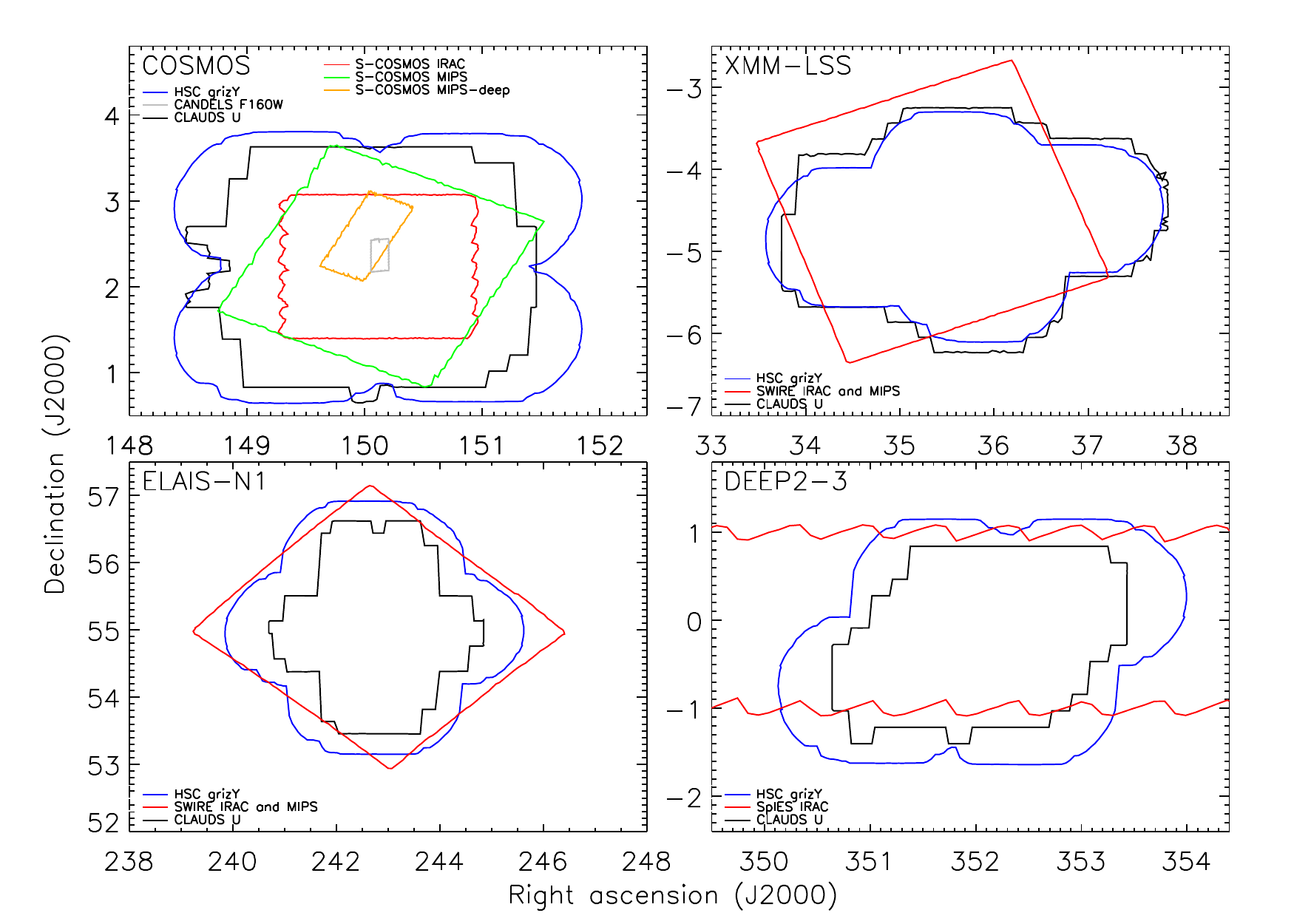}
\caption{
The multi-wavelength coverage in COSMOS, XMM-LSS, ELAIS-N1 and DEEP2-3 fields: 
HSC in blue; the Spitzer IRAC and MIPS in red. 
The ELAIS-N1 and XMM-LSS fields are covered by SWIRE; 
the DEEP2-3 field is covered by SpIES; 
and COSMOS field was covered by the S-COSMOS survey.
CFHT U-band coverage comes from the CLAUDS.
}\label{fig:coverage}
\end{figure}

The deep optical survey in the Hyper Suprime-Cam Subaru Strategic Program with Subaru telescope
was carried out in the \textit{grizY} bands\footnote{\url{https://smoka.nao.ac.jp/}}, 
covering four fields (Figure \ref{fig:coverage}): 
the Cosmological Evolution Survey \citep[COSMOS,][]{2007ApJS..172..150S}, 
the European Large-Area $ISO$ Survey-North 1 \citep[ELAIS-N1,][]{2000MNRAS.316..749O}, 
the XMM-Newton Large-Scale Structure \citep[XMM-LSS,][]{2004JCAP...09..011P} survey and 
the Deep Extragalactic Evolutionary Probe 2 field 3 \citep[DEEP2-3,][]{2013ApJS..208....5N}.
The total coverage is about 31 deg$^2$, the largest among existing deep surveys with comparable depths. 
The optical bands and their liming magnitudes are summarized in Table \ref{tab:field_coverage}. 
A matching U-band deep  survey was carried out in the CFHT Large Area U-band Deep Survey(CLAUDS)
 \citep{2019MNRAS.489.5202S}. 
CLAUDS was designed to match the HSC deep survey and made up what was missing in the original HSC survey. 
The HSC photometry was performed with fixed position and shapes for each detection, typically
in the i band\footnote{For details see \url{https://hsc-release.mtk.nao.ac.jp/doc/index.php/processing-2/}},
while the CFHT U-band data were reduced following the same process as HSC by the CLAUDS team.
Though in this work we aim to selected a bright sample, this deep survey permits the construction 
of an accurate SED for each object, and thus reliable photometric redshift and stellar mass measurements.

There are rich multi-wavelength ancillary data in these four fields. 
In particular, the $Spitzer$ IRAC and MIPS photometry data are essential in constraining their stellar masses, 
weeding out AGNs, and classifying their SED types. 
All four fields, except DEEP2-3 missing MIPS, 
were well covered by both IRAC and MIPS, as shown in Figure~\ref{fig:coverage},
and listed in Table~\ref{tab:field_coverage}.
The colors used in this paper are 
calculated with 2 arcsec apertures, while all single-band magnitudes are 
aperture corrected to the total flux.
The HSC data are corrected by the PSF provided by the HSC-SSP, 
while the \textit{Spitzer} data are corrected by the coefficients 
from S-COSMOS \citep{2007ApJS..172...86S} in the COSMOS fields, 
the \textit{Spitzer} Wide-area InfraRed Extragalactic Survey (SWIRE) Legacy Project \citep{2003PASP..115..897L}
in the ELAIS-N1 and XMM-LSS fields,
and the $Spitzer$ IRAC Equatorial Survey \citep[SpIES;][]{2016ApJS..225....1T} in the DEEP2-3 field.

\begin{deluxetable*}{lcccccccccc}
\tablenum{1}
\tablecaption{5$\sigma$ Limiting Magnitudes for CLAUDS, HSC and Spitzer Surveys}
\tablewidth{0pt}
\tablehead{
\colhead{Region name} &
\colhead{U} &
\colhead{g} &
\colhead{r} &
\colhead{i} &
\colhead{z} &
\colhead{Y} &
\colhead{$[$3.6$]$} &
\colhead{$[$4.5$]$} &
\colhead{$S_{24\mu m}$} \\
\colhead{ } & 
\colhead{(mag)} &
\colhead{(mag)} &
\colhead{(mag)} &
\colhead{(mag)} &
\colhead{(mag)} &
\colhead{(mag)} &
\colhead{(mag)} &
\colhead{(mag)} &
\colhead{(mJy)}\\
\colhead{(1)} & 
\colhead{(2)} &
\colhead{(3)} &
\colhead{(4)} &
\colhead{(5)} &
\colhead{(6)} &
\colhead{(7)} &
\colhead{(8)} &
\colhead{(9)} &
\colhead{(10)} 
}
\startdata
COSMOS & 27.7 & 27.3 & 27.2 & 26.7 & 26.3 & 26.6 & 23.7 & 24.5 & 0.3, 0.06$^{a}$\\
XMM-LSS & 27.7* & 26.8 & 26.5 & 26.6 & 25.4 & 25.3 & 21.9 & 21.9 & 0.3\\
ELAIS-N1 & 27.5 & 27.0 & 26.2 & 26.2 & 25.6 & 25.5 & 21.9 & 21.9 & 0.3\\
DEEP2-3 & 27.5 & 27.1 & 26.0 & 26.1 & 25.4 & 25.6 & 21.9 & 22.0 & -\\
\enddata
\tablecomments{
(1) Region name. 
(2)-(10)The 5$\sigma$ limiting magnitude in  CLAUDS U band (2), HSC grizY (3)-(7),
Spitzer IRAC (8)-(9) and MIPS (10).
*: U$_*$ used.}
\label{tab:field_coverage}
\end{deluxetable*}

\subsection{The Sample Selection} \label{sec:selection}

The Balmer break is often used to identified passive galaxies at various redshifts.
Yet single color may not be able to distinguish dusty and passive galaxies. 
One more parameter is needed to pin down the Balmer break. 
For example, using one color criterion and position information, 
one can identify clusters according to the morphology-density relation
 \citep{2005APJS.157.1G, 2007ASP.634..861Y, 2009ApJ.698.1943W}. 
Some studies even used an extreme NIR-MIR color to select passive galaxies at z$>$4
\citep{2011ApJ...742L..13H, 2012ApJ...750L..20C, 2019ApJ...876..135A}. 
In this paper, we apply a red $i - Y$ color and a luminosity prior to select passive galaxies at z$\sim$1.2.

\begin{figure}[ht!]
\includegraphics[width=0.45\textwidth]{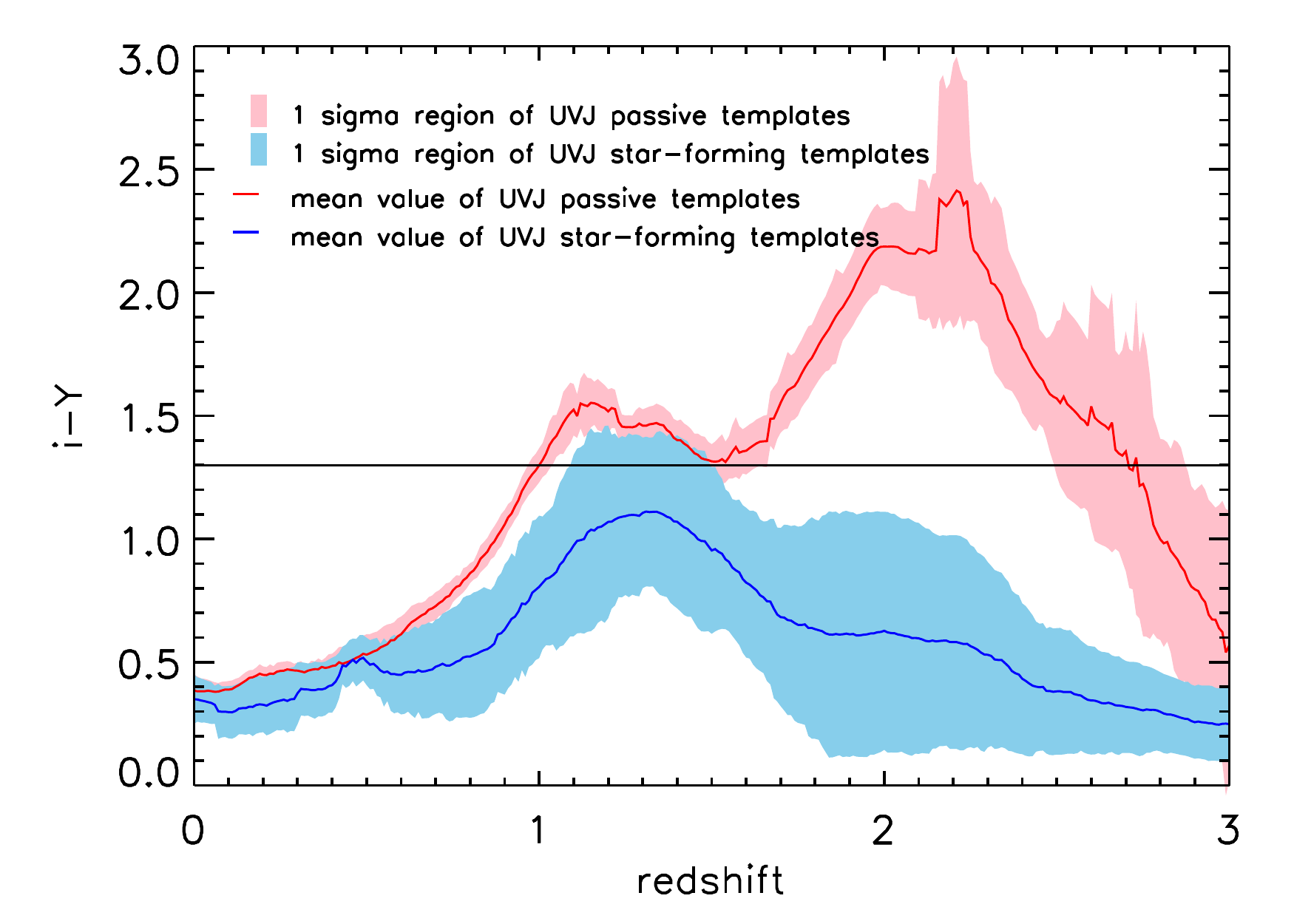}
\caption{The modeled color $i-Y$ $vs$ redshift using the galaxy templates from \citet{2014ApJS..212...18B}.  
The pink region shows the passive galaxy template distribution,
and red line is the mean value of these passive templates. 
The cyan area is the region of star-forming galaxy template distribution,
and blue line is the mean value of these star-forming templates. 
}
\label{fig:selection}
\end{figure}

\begin{figure}[ht!]

\includegraphics[width=0.45\textwidth]{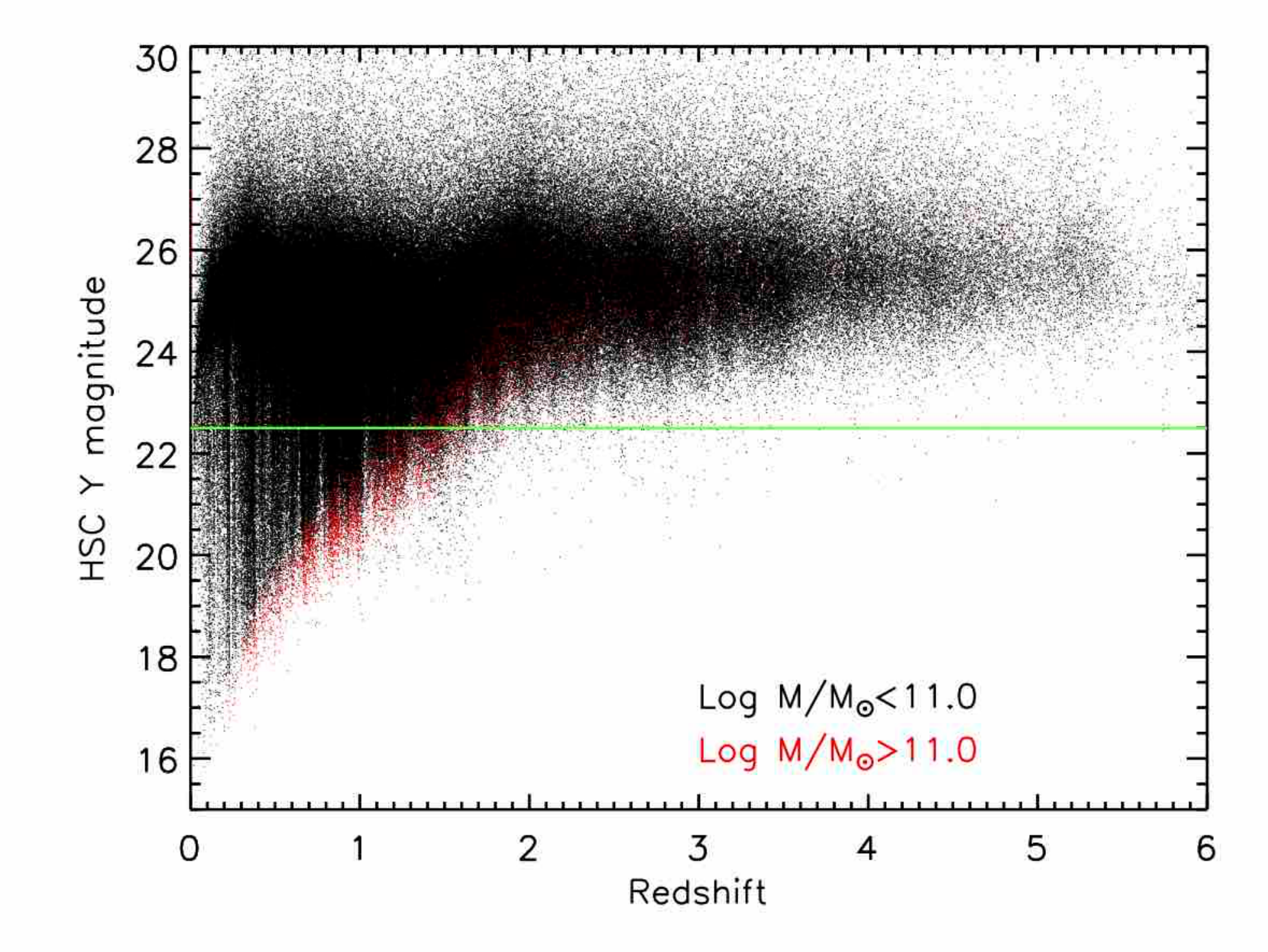}
\caption{HSC $Y$ $vs$ redshift from the SPLASH COSMOS catalog \citep{2016ApJS..224...24L}. 
Black points are galaxies with M$_*<10^{11.0}$$\msun$. 
Red points are those with M$_*>10^{11.0}$$\msun$. 
The green line is the $Y=22.5$ cut we selected to reject high $z$ sources. 
Very few galaxies at $z>1.5$ lie below $Y<22.5$.  
}
\label{fig:selectionone}
\end{figure}

\begin{figure}[ht!]

\includegraphics[width=0.45\textwidth]{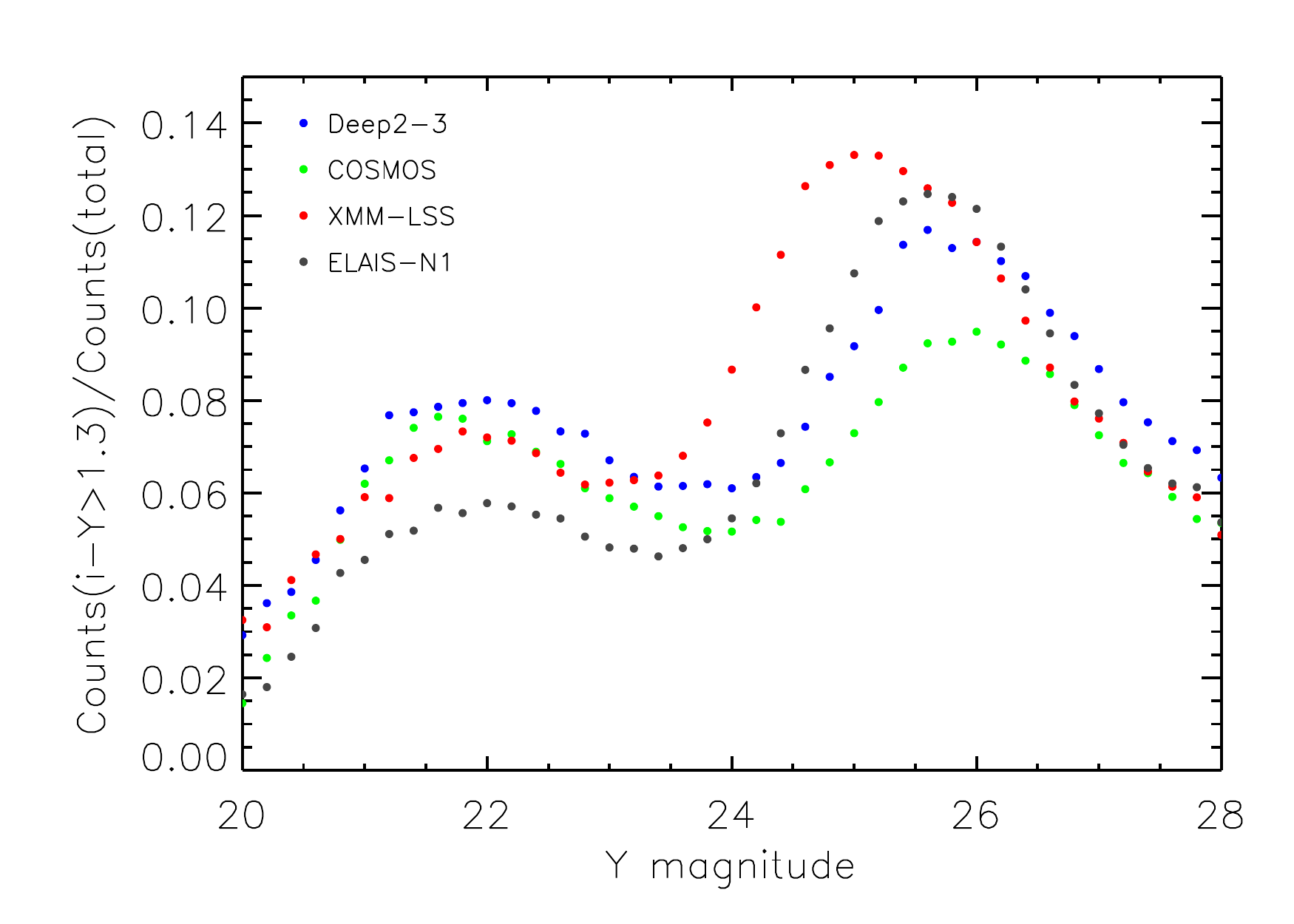}
\caption{The number count ratios of galaxies with i-Y$>$1.3 to the total number of galaxies in the four HSC fields. 
The two peaks correspond to the red galaxies at z$\sim$1.2 and z $\sim$ 2.2.}
\label{fig:countratio}
\end{figure}

At z$\sim$1.2, the Balmer break is redshifted to between the HSC $i$ and $Y$ band. 
All objects in our sample must have i band and Y band detections with a signal-to-noise ratio greater than 5.
We choose the $i-Y$ color instead of $z-Y$ as it better brackets the Balmer break. 
Figure \ref{fig:selection} shows the modeled  $i-Y$ colors as functions of redshifts. 
These modeled colors are derived using various galaxy SED templates from \citet{2014ApJS..212...18B}, 
which were generated from the observed spectral data of nearby galaxies.
The E(B-V) of these galaxy templates ranges from 0.01 to 0.37. 
This range is similar to the observed E(B-V) range at $1<z<1.5$,
as shown in the SPLASH survey \citep{2016ApJS..224...24L, 2018ApJS..235...36M}. 
We also explore the effect of reddening with our MIPS data in \S\ref{sec:uvj}. 
We then classify these SED templates in the $UVJ$
diagram \citet{2009ApJ...691.1879W}, and find 34 templates for passive galaxies. 
At z$>$1, all passive SED templates are at $i-Y>1.3$ (straight line in Figure~\ref{fig:selection}), 
while the remaining star forming templates 
predict that most have $i-Y<1.3$, regardless of redshifts. 
Thus we adopt the color cut as $i-Y=1.3$.
Figure \ref{fig:selection} also shows that this color criterion selects passive galaxies 
in a wide redshift range of 1$< z <$ 3. 
We then propose an additional luminosity cut to reject galaxies at higher redshifts.
In practice, we adopt a very bright limiting magnitude, Y=22.5 mag, to the sample selection. 
This criterion can reject almost all galaxies at $z>1.5$, as shown in Figure \ref{fig:selectionone}. 
with a complete Y selected sample in the COSMOS field \citep{2016ApJS..224...24L}. 
This can be more clearly illustrated in Figure \ref{fig:countratio}, in which
we plot the ratio of number counts for galaxies with $i-Y>1.3$ to the total galaxy number counts as  function of Y band magnitude. 
The two peaks in this plot corresponding to the two peaks in Figure~\ref{fig:selection}. 
The first peak in Figure~\ref{fig:countratio} is for red galaxies at z$\sim$1.2, which drops at Y$\sim$22.5. 
This justifies our bright limiting magnitude choice for sample selection. 
So our selection criteria for passive galaxies at z$\sim$1.2 are: 
\begin{equation}
i-Y>1.3\  \&\  Y<22.5
\end{equation}
To further test this selection,
we generate a mock catalog using the galaxy luminosity functions {\citep{2005ApJ...634..861Y}}
in 1$<z<$2 and real galaxy SED templates. 
After applying both criteria to the mock catalog, we found a low contamination of  z$>$1.5 objects
of only 3.1\%.

\begin{figure}
\includegraphics[width=0.45\textwidth]{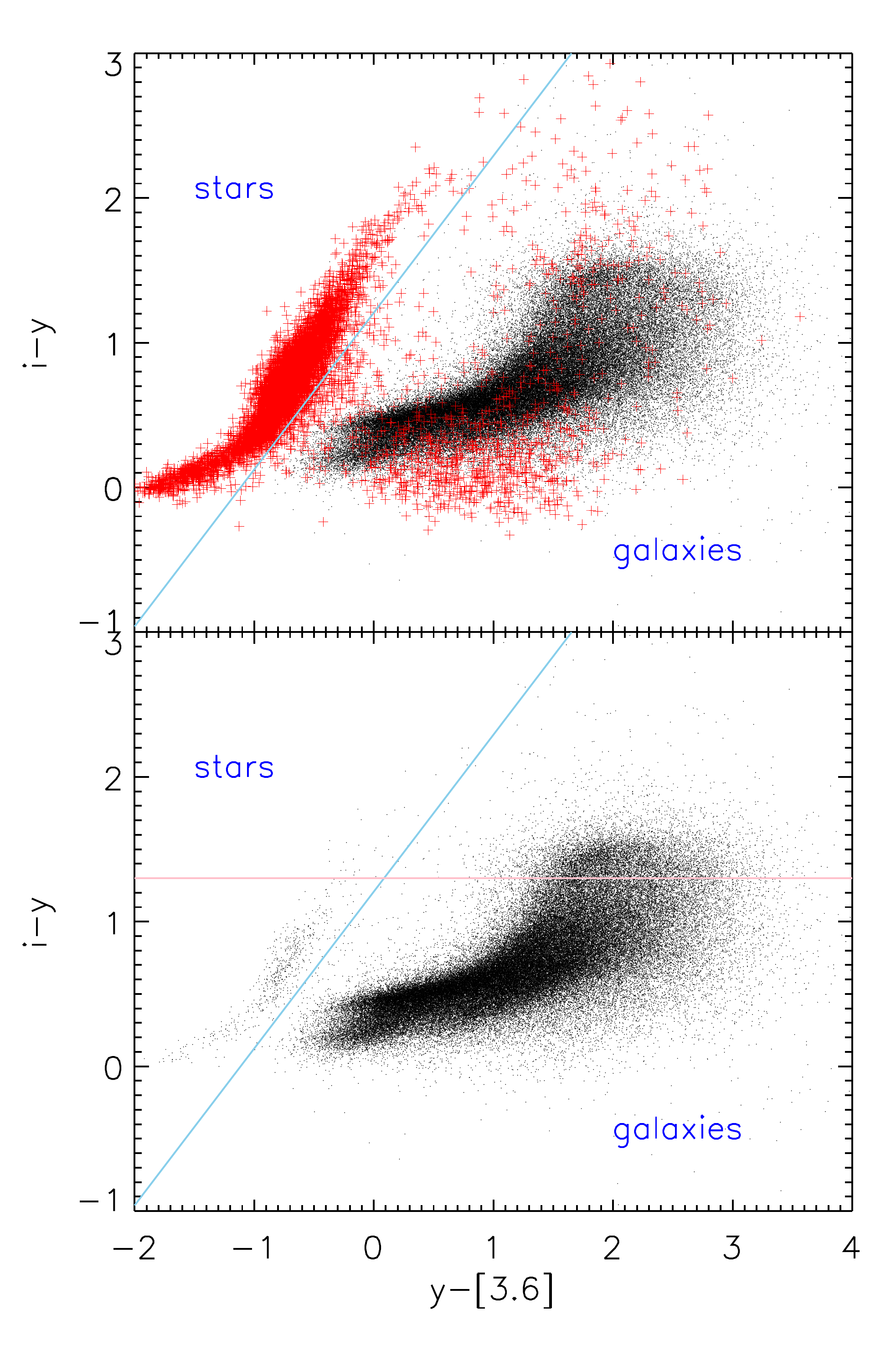}
\caption{The $i-Y$ $vs$ $Y-[3.6]$ color-color diagram. 
Stars and galaxies are well separated by the skyblue line from i-y=1.08*(y-[3.6])+1.21. 
{(Top)} The red crosses are objects with Y band extendedness parameter of 0 (i.e. stars). 
The black points are objects with extendedness parameter of 1 (i.e. galaxies).
{(Bottom)} Objects remaining after removing the stars only by their HSC extendedness. 
After applying the i-Y color selection in our  sample (pink line), 
we expect the star contamination to be less than 0.5\% in our full sample. 
}
\label{fig:sg2}
\end{figure}

Our sample is then selected from the deep HSC catalog, 
which has a pixel scale $=$ 0.168''/pixel and an average seeing of $\sim$0.8''
in the Subaru Y band.
We first apply the above criteria to the parent catalogs in the fours fields.
Then we reject stars in the selected sample by 
their $i - Y$ vs $Y - [3.6]$ colors, similar to \citet{2013ApJ...766...21H}. 
The star-galaxy separation line is defined in Figure \ref{fig:sg2} (skyblue line). 
After that, we check the possible contamination and completeness of our sample
with their HSC Y band extendedness parameters\footnote{\url{http://hsc-release.mtk.nao.ac.jp/doc/index.php/stargalaxy-separation/}}. 
Since with our bright limiting magnitude (Y$<$22.5), 
the extendedness measurement is very accurate. 
Figure \ref{fig:sg2} shows that the star-galaxy separation by extendedness 
and by color-color classification are consistent with each other.
With our selection ($i - Y > 1.3$), 
only 2\% of the objects that fall in the galaxy region 
in the $i - Y$ vs $Y - [3.6]$ diagram (Figure \ref{fig:sg2} bottom panel, upper right region)
have an extended parameter of 0 (for point source). 
We suspect these are likely compact galaxies at higher redshift that cannot be 
resolved, thus we keep them in the final sample. 
In the case that these point-source are indeed stars,
this 2\% marks the upper limit for the star contamination in our sample. 
On the other hand, in the HSC regions with no IRAC coverage (1/3 of our sample), 
we apply the extendedness parameter to reject stars.
In these regions, we may miss up to 2\% galaxies which are compact
due to the lack of IRAC data.
If these objects are indeed stars,
the star contamination in our final sample is expected to be less than 1.3\%. 

Our final sample consists of a total of 33,893 objects. 
Ancillary data, in particular the MIR data, are essential in determining their redshifts, stellar mass, and AGN fractions. 
Due to $Y-[3.6]$ cut as shown in Figure \ref{fig:sg2} (skyblue line), 
we expect that almost all objects with $Y< 22.5$ in the sample are brighter than the $[3.6]_{limit}=21.9$. 
In fact, in the IRAC covered area, we found an IRAC counterpart for 95\% of the selected objects.
Objects not detected in 3.6\,\micron\ are mostly on the edge of the coverage map,
due to the relatively shallower depth.
The MIPS 24\,\micron\ coverage is about the same as the IRAC. 
Yet there are only 264 objects detected at 24\,\micron, $\sim$1\% of the whole sample. 
We will discuss these objects in \S\ref{sec:uvj}. 
Table \ref{tab:field} summarizes the number counts in our selected sample
after masking out the bright star region files 
from HSC\footnote{\url{https://hsc-release.mtk.nao.ac.jp/doc/index.php/bright-star-masks-2/}}.
To achieve accurate photometric redshift and stellar mass for validation purpose, 
all photometry data are corrected to the total flux with relevant point spread function (PSF).
All statistic properties in \S\ref{sec:validation} are calculated excluding the 24$\mu$m data.

\begin{deluxetable*}{lccccccccccc}
\tablenum{2}
\tablecaption{Number counts in different fields} 
\tablewidth{0pt}
\tablehead{
\colhead{Field name} &
\colhead{Area$_{total}$} &
\colhead{N$_{total}$} &
\colhead{N$_{star}$} &
\colhead{N$_{galaxy}$} &
\colhead{Area$_{IRAC}$} &
\colhead{N$_{galaxy}$} &
\colhead{N$_{24\mu m}$} &\\
\colhead{ } & 
\colhead{($deg^{2}$)} &
\colhead{$^{Y<22.5}_{i-Y>1.3}$} &
\colhead{ } &
\colhead{ } & 
\colhead{($deg^{2}$)} & 
\colhead{ } & 
\colhead{ } &\\
\colhead{(1)} & 
\colhead{(2)} &
\colhead{(3)} &
\colhead{(4)} &
\colhead{(5)} &
\colhead{(6)} &
\colhead{(7)} &
\colhead{(8)} &
}
\startdata
COSMOS & 8.7 & 11719 & 1440 & 11867 & 2.2 & 3014 & 26, 26*\\
XMM-LSS & 7.3 & 9509 & 875 & 8635 & 5.3 & 6459 & 133 \\
ELAIS-N1 & 7.3 & 7533 & 1094 & 6440 & 6.1 & 5645 & 105 \\
DEEP2-3 & 7.9 & 9642 & 1105 & 8538 & 5.1 & 6014 & - \\
\hline
Total & 31.2 & 38403 & 4510 & 33893 & 18.7 & 21132 & 264 \\
\enddata
\tablecomments{
(1) The regions name.   
(2) The HSC coverage in each field.
(3) The number of objects satisfying the sample selection.
(4) The star selected by HSC Y band extendedness equal to 0.  
(5) The galaxy selected by HSC Y band extendedness equal to 1.
(6) The overlapping area of HSC and IRAC in each field.
(7) Number of galaxies in the overlapping area in (6).
(8) The number of galaxies with 24$\mu$m detection.
*: For the COSMOS 24$\mu$m deep field, 
an additional 26 objects are found to be above the 0.06mJy detection limit but $<$0.3 mJy.}
\label{tab:field}
\end{deluxetable*}

\section{Physical Properties of the Sample} \label{sec:validation}

\subsection{Photometric Redshifts Validation} \label{sec:photoz}

\begin{figure}
\includegraphics[width=0.45\textwidth]{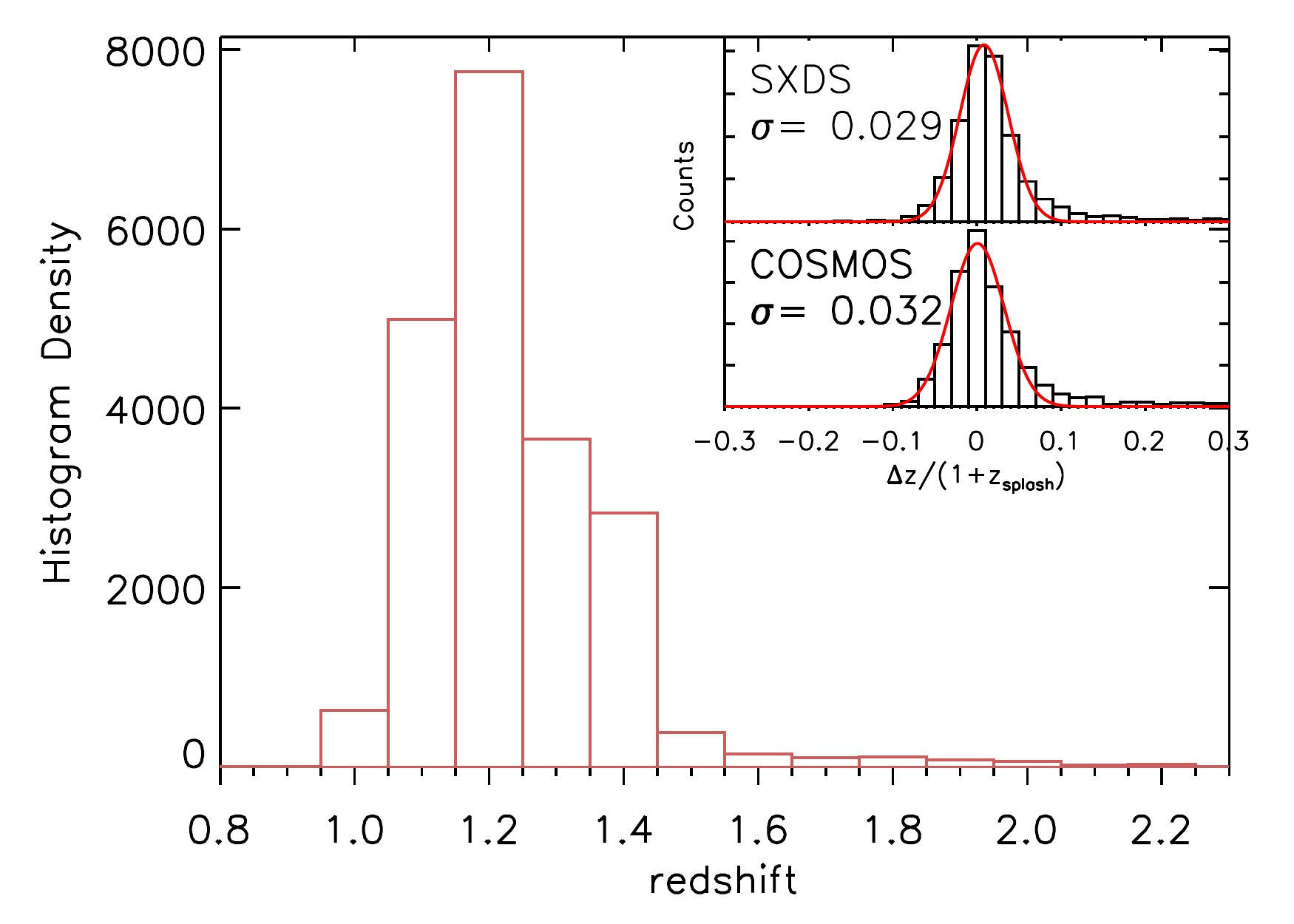}
\caption{
Histogram of the photometric redshift distribution of our sample, 
estimated from EAZY \citep{2008ApJ...686.1503B} with the 
CLAUDS U, HSC grizY, and IRAC 3.6 \& 4.5 \micron\ data.
The majority ($\sim$97\%) of our sample lies between $0.9<z<1.5$, with a peak of z$=$1.2.  
Up-right corner shows the photometric redshift comparison between our sample and the archive values using SPLASH \citep{2016ApJS..224...24L, 2018ApJS..235...36M}.
In the COSMOS and SXDS fields, 
the redshift comparison yields $\sigma$($\Delta$z/1+z) of $\sim$0.03. 
Considering the intrinsic $\sigma$($\Delta$z/1+z)$\sim$0.02 of the SPLASH redshifts, 
this yields $\sigma$($\Delta$z/1+z)$\sim$0.03 for our photometric redshifts.
}
\label{fig:photoz}
\end{figure}

Redshifts are essential in deriving the physical properties for galaxies.  
Since spectroscopy for passive galaxies at $z>1$ is difficult and expensive, 
we rely on the photometric redshifts for our sample. 
Though photometric redshifts are already available in the COSMOS and part of the XMM fields
 \citep{2016ApJS..224...24L, 2018ApJS..235...36M}, 
to be consistent, we derive the photometric redshifts for the whole sample with the same method.  
Although IRAC data are only available for about 2/3 of the covered area,
we argue that IRAC photometry is required to ensure accurate photometric redshifts for this sample. 
Therefore, 
photometric redshifts (Figure~\ref{fig:photoz}) were estimated only for galaxies in the IRAC covered areas
using the EAZY code \citep{2008ApJ...686.1503B},
based on the CLAUDS $U$, HSC $grizY$ and IRAC 3.6 \&\ 4.5 photometries. 
Figure \ref{fig:photoz} shows that 
the majority (96.5\%) of our sample are in 1.0 $ < z < $ 1.5, with a peak of z=1.2.
There are only 0.2\% galaxies at $z<1.0$, 
and 3.3\% at $z>1.5$. 
This is consistent with our model prediction. 
We also compare our photometric redshifts with those available in the COSMOS and XMM fields
\citep{2016ApJS..224...24L, 2018ApJS..235...36M}. 
Compared to the existing photo-$z$,
the accuracy of our photometric redshifts is estimated to be $\sigma$($\Delta$z/1+z) of $\sim$0.03.

\begin{deluxetable*}{ccl}
\tablenum{3}
\tablecaption{Catalog Structure for the Massive Passive Galaxy Sample} 
\label{tab:cat}
\tablewidth{0pt}
\tablehead{
\colhead{(1)} & 
\colhead{(2)} &
\colhead{(3)} 
}
\startdata
\hline
ID &   & Object ID in HSC-SSP pdr2 catalog \\
RA & $deg$ & Right Ascension, decimal degreee (J2000)  \\
DEC & $deg$ & Declination, decimal degreee (J2000)  \\
$i-Y$ & mag & Hyper-Supreme-Cam $i-Y$ color   \\
$Ymag$ & mag & Hyper-Supreme-Cam $Y$ band magnitude \\
zphot &   & Photometric redshift by SED fitting using the EAZY code \\
logMstar &  $log(\msun)$  & Stellar mass by SED fitting using the FAST code\\
Field &   & Field name \\
\enddata
\end{deluxetable*}

\subsection{SED Classification} \label{sec:uvj}

The rest-frame color-color diagram is a straightforward way to separate passive and dusty galaxies. 
In the rest-frame $UVJ$ diagram, the direction of dust reddening is rather different from that of the stellar population. 
As a result, passive galaxies locate in a separate region from the dusty galaxies\citep{2005ApJ...624L..81L,2009ApJ...691.1879W}.
The rest-frame \textit{U}, \textit{V} and \textit{J} are closest to 
the observed \textit{z}, \textit{Y} and IRAC 3.6 $\mu$m bands for this sample at $z\sim1.2$,
so we calculated the rest-frame colors accordingly with K-correction. 

\begin{figure}
\includegraphics[width=0.45\textwidth]{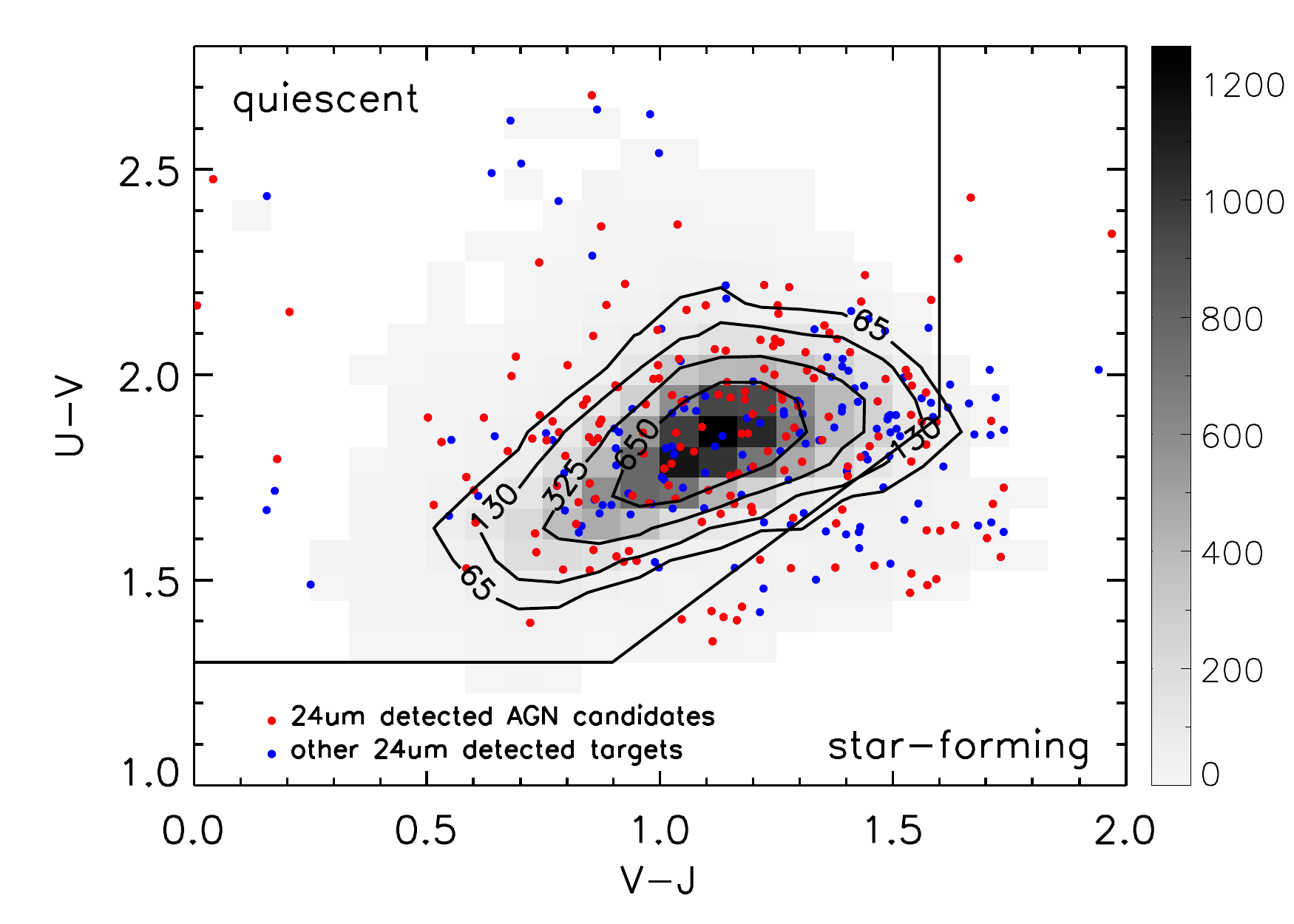}
\caption{The rest-frame $UVJ$ diagram. 
The grey scales represents the density of all objects in our sample.
The contours represent 50\%, 25\%, 10\%, and 5\% of the maximum density, respectively.
The red dots are the 24$\micron$ sources identified as AGNs (See \S\ref{sec:uvj} and Figure~\ref{fig:agn}). 
About {95\%} of our objects are in the $UVJ$ quiescent galaxy region. 
We adopt the rest-frame \textit{U}, \textit{V} from \citet{1990PASP..102.1181B} and \textit{J} from \citet{2002PASP..114..180T}.
The K-corrections are carried out during the SED fitting procedure.}
\label{fig:uvj}
\end{figure}

Most galaxies in our sample, $\sim$95\% 
fall in the $UVJ$ quiescent galaxy region in Figure \ref{fig:uvj}, confirming their passive nature. 
Only $\sim$5\% of our sample are in the area of star-forming, dusty galaxy region. 
Galaxies of our sample in the COSMOS field have counterparts in a deep multi-wavelength catalog, 
in which each galaxy SED was already classified. 
About 78\% of our galaxies in COSMOS field were identified as passive galaxies this way \citep{2016ApJS..224...24L}. 
Both data sets confirm a consistent result that most galaxies in our sample are passive galaxies.

\begin{figure}
\includegraphics[width=0.45\textwidth]{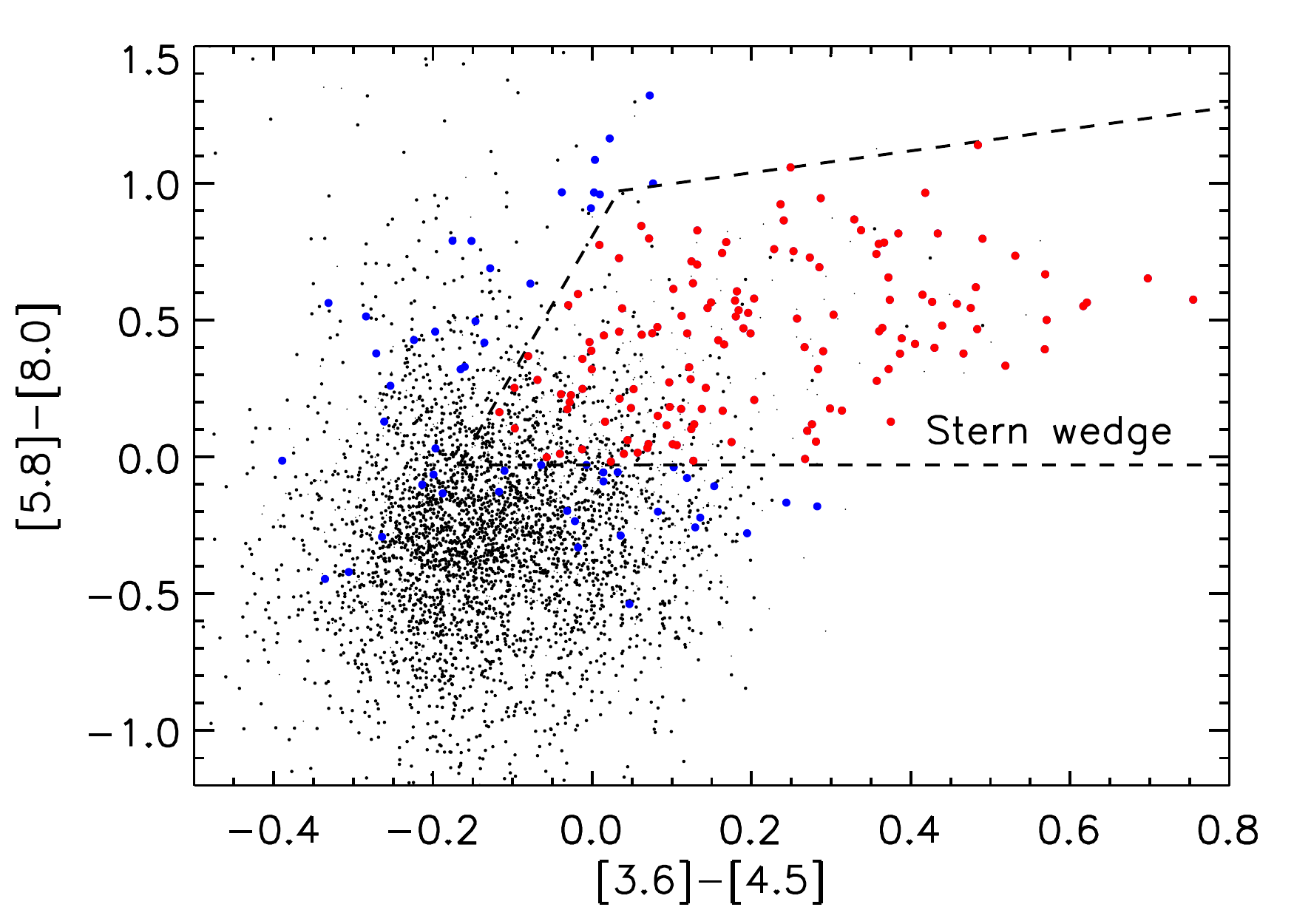}
\includegraphics[width=0.45\textwidth]{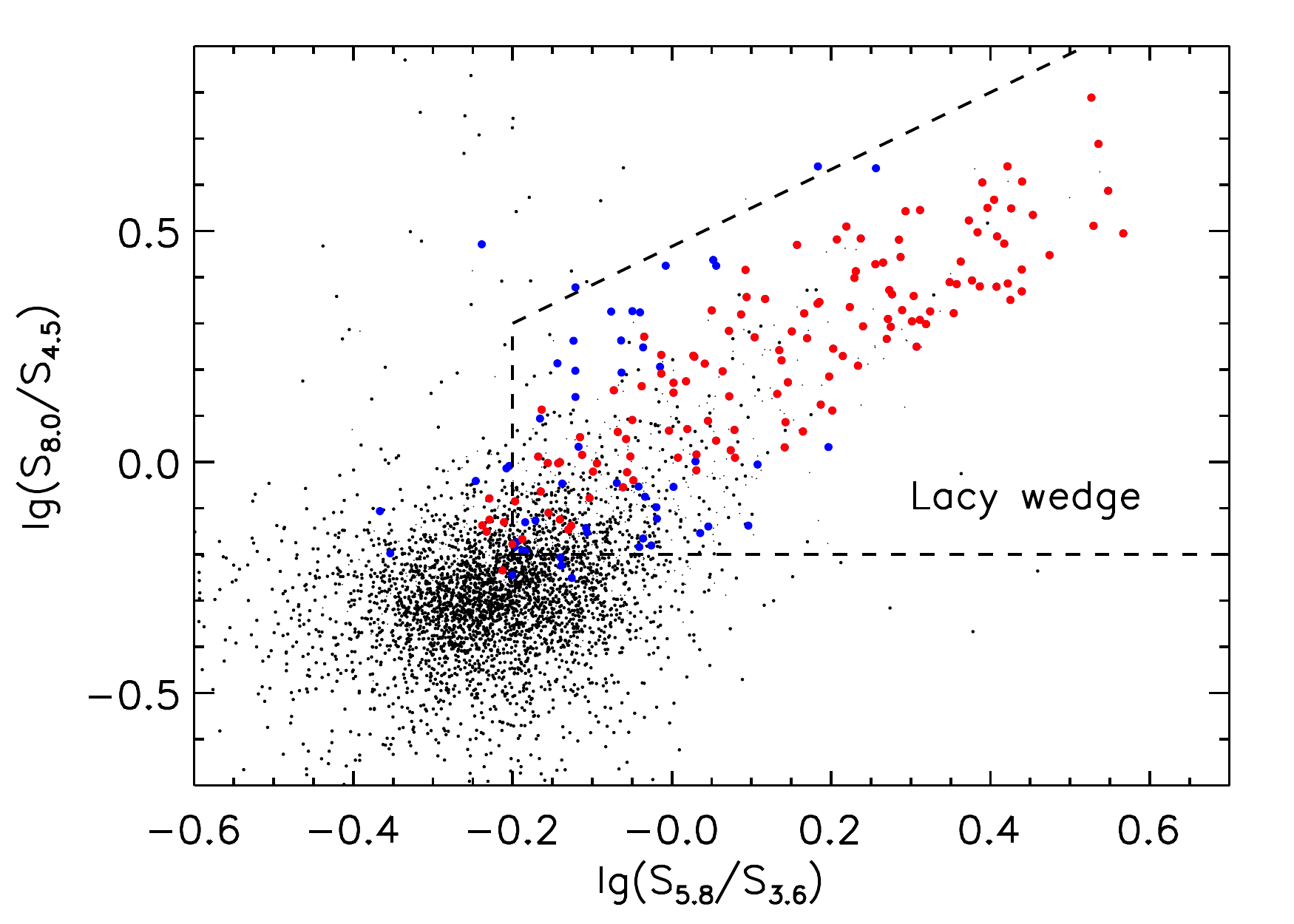}
\caption{
{\it Top}: IRAC [3.6]-[4.5] $vs$ [5.8]-[8.0] color diagram (Stern Wedge, \citet{2005ApJ...631..163S}). 
About 70\% of the 24\,\micron\ sources are identified 
in the AGNs (red) area.
The  24\,\micron\ sources outside of the Stern wedge are marked as blue points.
{\it Bottom}: IRAC lg($S_{5.8}/S_{4.5}$) $vs$ lg($S_{8.0}/S_{4.5}$) diagram (Lacy wedge, \citet{2004ApJS..154..166L}). 
About 88\% of the 24\,\micron\ sources are identified 
in the AGNs (red) area. 
To compare the two wedge selections, the color codes in the right panel
is the same as in the left panel. 
Most of the 24$\mu$ sources outside of the Stern wedge (blue points) are also identified as AGNs in the Lacy Wedge. 
Combining the two color selections, a total of 92\% of the 24\,\micron\ sources (with IRAC data)
are in the AGNs (red) area. } 
\label{fig:agn}
\end{figure}

As mentioned earlier, there are 264 galaxies in our sample detected at 24\,\micron. 
Normally galaxies with MIPS 24\,\micron\ detections are dusty galaxies. 
The MIPS 24\,\micron\ limiting flux densities are 0.3 mJy in the SWIRE covered area and 0.06 mJy in the deep area in COSMOS.
These two limiting flux densities corresponds to star formation rate (SFR) of 20-50\,$\msun$/yr, 
slightly above the main sequence value at z$\sim$1.2 \citep{2014ApJS..214...15S} 
for galaxies in the same mass range of our sample. 
Dusty star forming galaxies on or below the main sequence may not be detected at 24\,\micron. 
Once detected, they are often qualified as luminous Infrared Galaxies (LIRGs) 
if the 24\,\micron\ emission is purely powered by star formation activities. 
Yet, most of the 24\,\micron-detected galaxies in our sample are located in the passive galaxy region in the $UVJ$ diagram.
This in turn suggests an AGN origin of the MIPS 24\,\micron\ emission.  
AGN can also contribute excessive emission in the IRAC bands for these galaxies.
We plot our sample in the IRAC color-color diagrams for AGN selections in Figure \ref{fig:agn}. 
Of the 24\,\micron\ detected galaxies that have all 4 IRAC bands,
92 \% qualify as MIR-selected AGNs according to the criteria in 
either \citet{2004ApJS..154..166L} or  \citet{2005ApJ...631..163S}
In further analysis, we exclude all MIPS 24 sources to avoid contamination from AGNs.

\subsection{Morphological Properties}\label{sec:mor}

\begin{figure*}[ht]
\subfigure{
\includegraphics[width=1\textwidth]{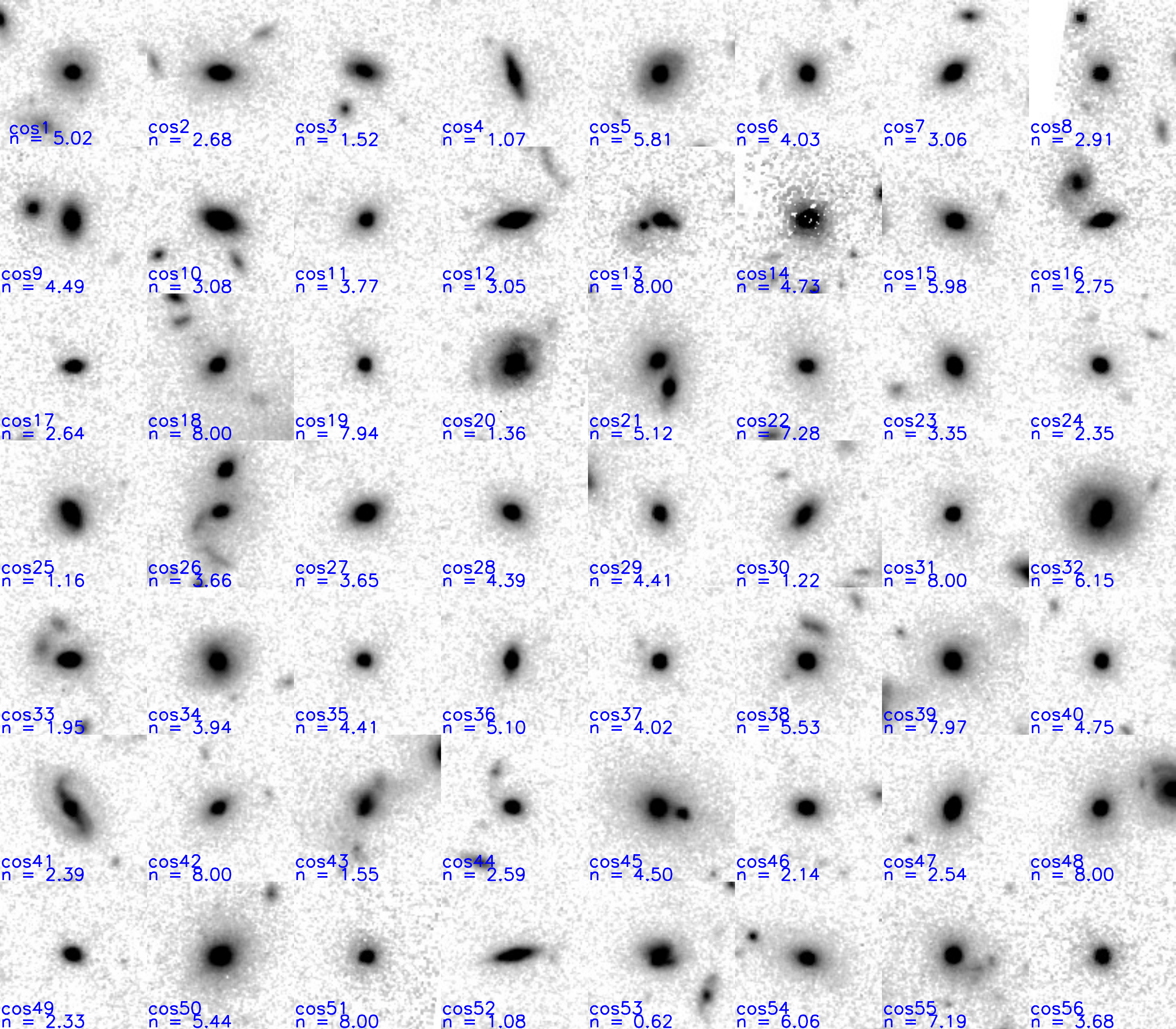}
}
\caption{HST F160W images from CANDELS COSMOS field. 
Each stamp has a size of 6"x6", and the fitted s\'{e}rsic index is marked for each source.
Most of the 116 galaxies in our sample covered by CANDELS
are visually inspected and identified as E/S0 type galaxies.
For more details of the morphological classifications see \S\ref{sec:mor}.
COS9,20,45 and 53 are found to have nearby objects with $\Delta$z$\sim$0.03.} 
\label{fig:img1}
\end{figure*}

\begin{figure*}[ht]
\includegraphics[width=1\textwidth]{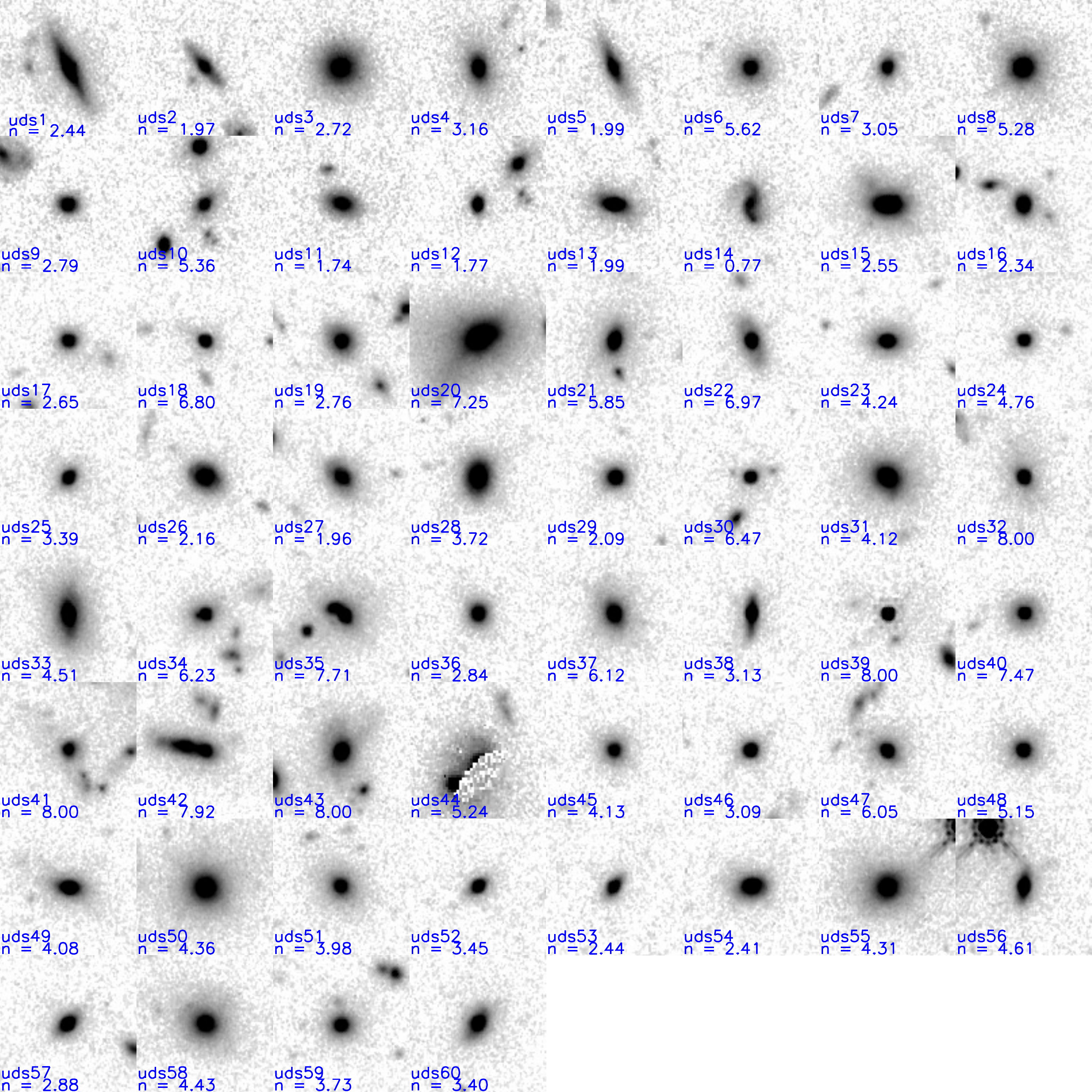}
\caption{HST F160W images from CANDELS UDS field. 
Each stamp has a size of 6"x6", and the fitted s\'{e}rsic index is marked for each source.
Most of the 116 galaxies in our sample covered by CANDELS
are visually inspected and identified as E/S0 type galaxies. 
For more details of the morphological classifications see \S\ref{sec:mor}. 
UDS44 is heavily contaminated by bad pixels in the HST image.
UDS1,2,7,21,41 and 43 are found to have nearby objects with $\Delta$z$\sim$0.03.
} 
\label{fig:img2}
\end{figure*}

In this section we present the morphological properties of our sample,
both by visual classification and by morphological parameters.
The sample was mainly selected in the Subaru Y-band. 
Even with the super seeing condition of 0.8" in the Subaru imaging, 
this still limits our resolution to $\sim$7kpc at z=1.2 in physical scale. 
Only large galaxies can be resolved in the Subaru Y-band image, 
which is not representative of our sample. 
Fortunately, there are two deep HST coverages among our fields 
that were observed in the Cosmic Assembly Near-Infrared Deep Extragalactic Legacy Survey (CANDELS) 
\citep{2011ApJS..197...35G, 2011ApJS..197...36K}. 
A total of 116 galaxies in our sample are in the CANDELS covered areas. 
Morphologies of all 116 galaxies were visually identified as 
E/S0 galaxies (107), 
Spiral galaxies (6) and others (3)
by our team members. 
We show the morphologies of these 116 galaxies in Figure \ref{fig:img1} \& \ref{fig:img2}.
A total of 10 galaxies have nearby objects 
with a $\Delta$z$\sim$0.03, based on photoz from CANDELS.

\begin{figure}
\includegraphics[width=0.45\textwidth]{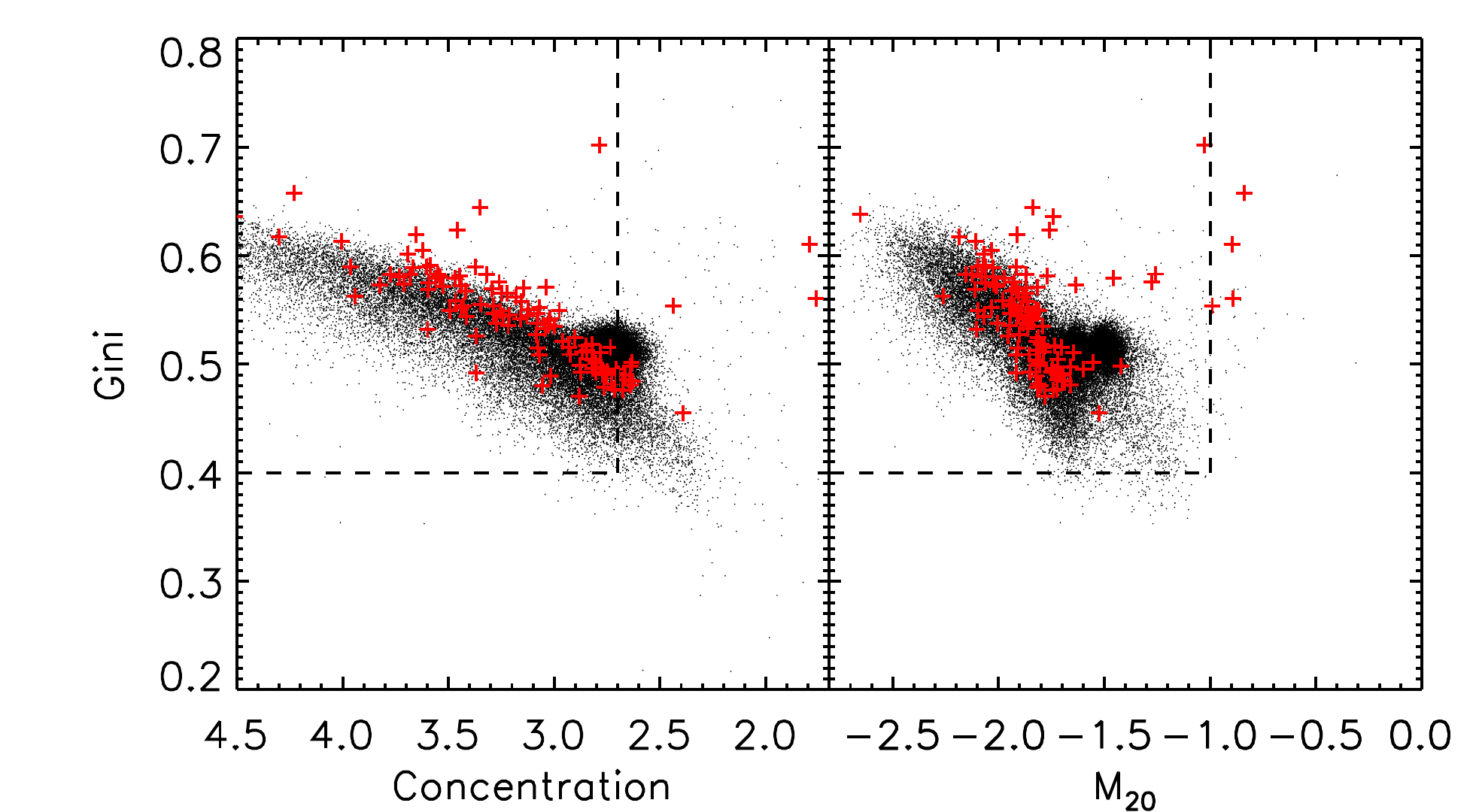}
\caption{Gini $vs$ M20 and Gini $vs$ C distribution of passive galaxies. 
The red cross represents our subsample of 116 galaxies covered by CANDELS. 
The black points mark the quiescent galaxies from \citet{2007ApJS..172..270C}. 
}
\label{fig:gini_m20}
\end{figure}

We also measured the morphological parameters. 
We adopted the Gini-M20 and Gini-C methods from
\citet{2003ApJ...588..218A, 2004AJ....128..163L,2008ApJ...672..177L}. 
Here C is the concentration parameter \citep{2000AJ....119.2645B}. 
\citet{2007ApJS..172..270C} found that passive galaxies
locate in separate areas from other types of galaxies in both Gini-M20 and Gini-C diagrams. 
We plot the Gini, M20, and C parameters for our 116 galaxies in Figure \ref{fig:gini_m20}.
The result of our morphological analysis shows that 
the majority of our galaxies (86\%) have early-type morphologies \citep{2007ApJS..172..270C}.

\begin{figure}
\includegraphics[width=0.45\textwidth]{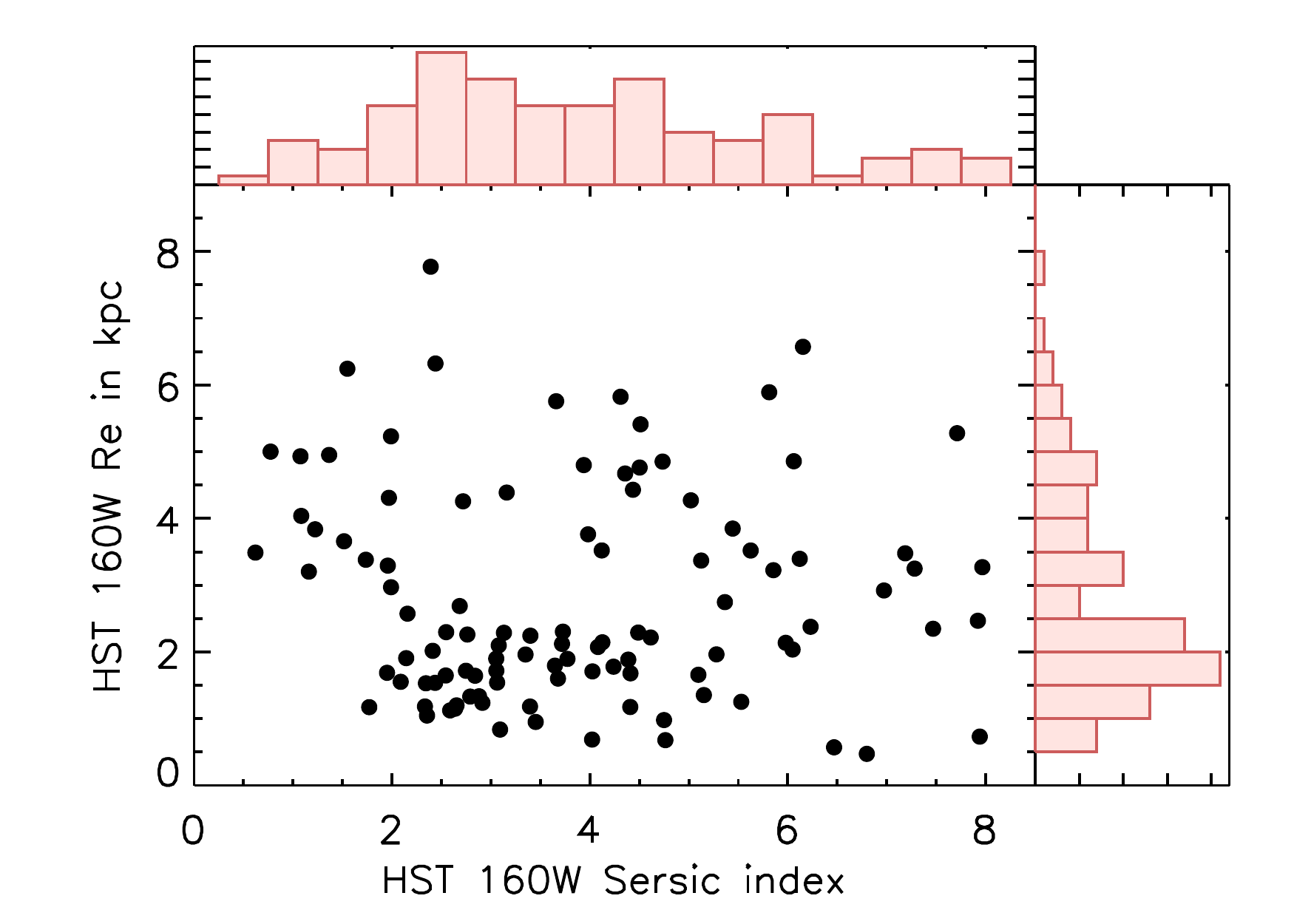}
\caption{
The effective radius $R_e$ versus the S\'{e}rsic index for our subsample in the CANDELS fields. 
Both values were derived with GALFIT for CANDELS galaxies from \citet{2012ApJS..203...24V}. 
Most of our galaxies have S\'{e}sic indices of 2$<n<$5.
 The peak of $R_e$ is at $R_e$=1.7 kpc,
 indicating that our sample is dominated by compact galaxies.
}
\label{fig:sersic}
\end{figure}

Galaxy surface brightness profile is another commonly used morphological classifier:
elliptical galaxies usually have high s\'{e}rsic indices (n) of n$\sim$4, 
while disk galaxies show exponential profiles. 
We take advantages of the CANDELS GALFIT measurements for these 116 galaxies \citep{2012ApJS..203...24V}. 
Most galaxies (91\%) in our sample have S\'{e}rsic index in the range of 2$<n<$5 (Figure~\ref{fig:sersic}), 
indicating bulge dominant or elliptical galaxy profiles. 
There are only 6\% of our galaxies with a S\'{e}rsic index of $n<1.5$,
indicating disk-dominant profiles.
Figure \ref{fig:sersic} also shows that
the effective radius ($R_e$) of most galaxies are between 1 kpc$<R_e<$3 kpc, with a peak $R_e\sim$1.75 kpc.
There are only 4 galaxies  with a larger size up to $R_e$=6$\sim$8 kpc.
Our distribution is similar to the morphologically selected elliptical galaxies 
at z=1.2 in the CANDELS fields \citep{2017A&A...597A.122S}.

In short, morphological studies of this HST subsample show that 
our color selection yields mostly morphological classified bulge dominant or elliptical galaxies. 
Since this subsample was selected in the same way as the whole
sample, this implies that our whole sample likely has the same
morphological types, dominated by bulge and elliptical galaxies.
This is expected for a sample of galaxies dominated by passive objects.

\subsection{Stellar Mass and Mass Function}\label{sec:mf}

\begin{figure}
\includegraphics[width=0.45\textwidth]{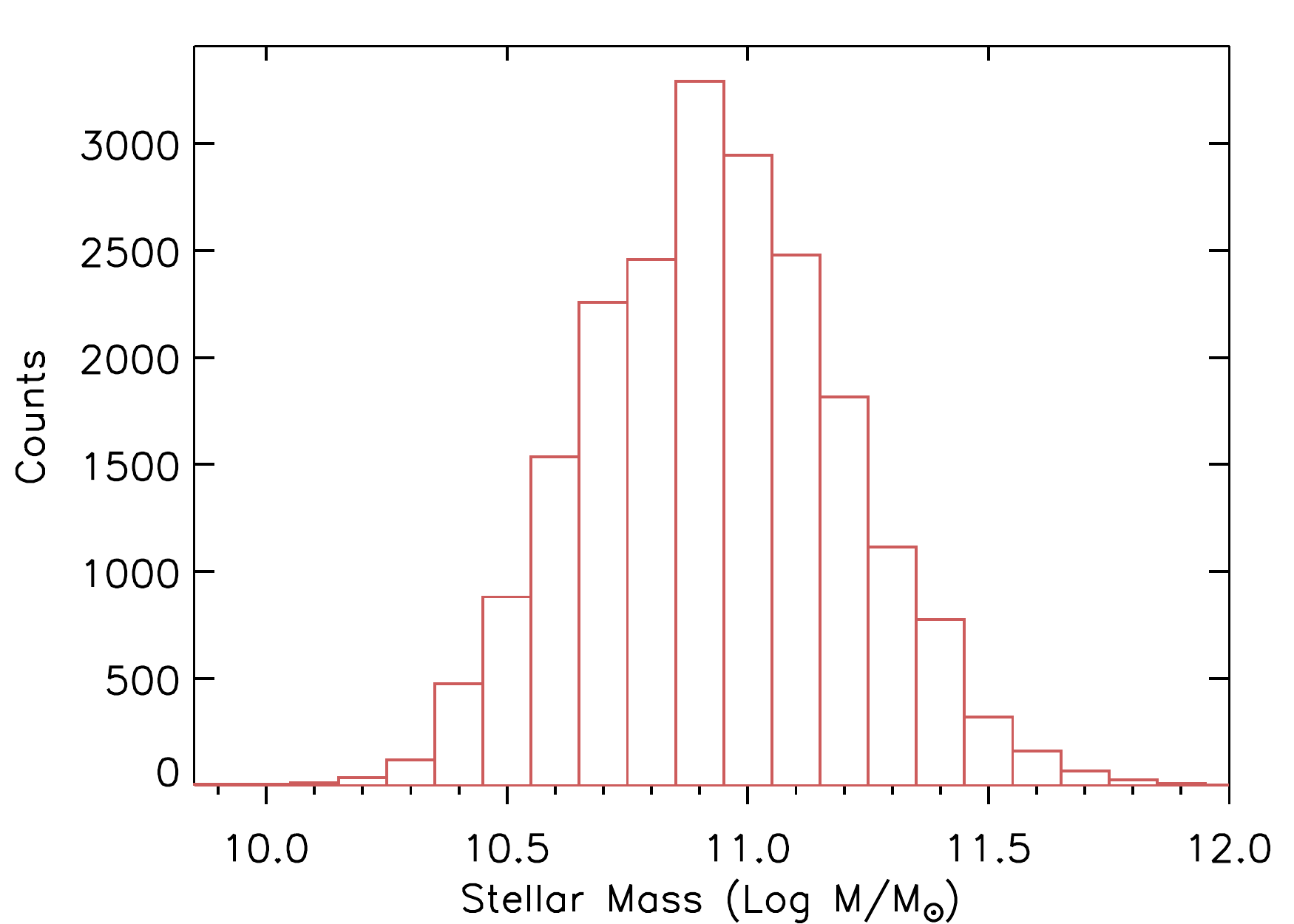}
\caption{The stellar mass distribution of our sample. 
The majority (95\%) of our galaxies are massive,
with a mass range of M$_{*}$ $>$10$^{10.5}\msun$,
with a peak at 10$^{10.9}\msun$, and a mean and median of 10$^{11.0}\msun$. 
} 
\label{fig:mass}
\end{figure}

Stellar masses for objects in our sample are estimated from the SED,
with CLAUDS $U$, HSC $grizY$, and IRAC 3.6 and 4.5\,\micron\ photometries. 
We only derived masses for galaxies with IRAC coverage (i.e. $\sim$2/3 of the whole area), 
using the FAST code \citep{2009ApJ...700..221K} with BC03 templates \citep{2003MNRAS.344.1000B}.
We obtained the stellar masses for 21,132 galaxies, roughly 62\% of the sample. 
The stellar mass of our sample is in range of $10.2<$log(M$_*/\msun)<12$ (Figure \ref{fig:mass}), 
yet about half have log(M$_*/\msun)>11$. 
There is a Spitzer warm mission survey covering the remaining area in the HSC fields (PI: Sajina). 
The final catalog for this survey is not available yet. 
We will derive stellar masses for remaining galaxies in the sample when the new IRAC data are available. 
Even for the subsample with stellar mass, 
it is already the largest among existing passive galaxy samples in the similar redshift range (1 $< z <$ 1.5).
We provide the passive galaxy sample with stellar mass estimate in a machine-readable form 
in the online journal, and the parameters provided in the catalog is listed in 
Table~\ref{tab:cat}.

In previous surveys, the largest sample for passive galaxies in similar redshift range was from the COSMOS field.
\citet{2013A&A...556A..55I} used a K-band selected sample to classify star forming and passive galaxies 
in 0.2$<z<$4, and obtained 3769 passive galaxies in $1.1<z<1.5$. 
The UltraVISTA survey also performed a K-band survey in COSMOS field, 
and obtained 6455 passive galaxies in $1.1<z<1.5$ \citep{2013ApJ...777...18M}. 
Our sample is a factor of 5-10 $\times$ larger than number of galaxies in both samples. 
Yet both surveys are much deeper than our survey, so they are more complete at the lower mass end.
Our sample is complete at log(M$_*/\msun)>10.5$, as shown in Figure \ref{fig:mass}. 
In this mass range, our sample is a factor of 7--20 $\times$ 
larger\footnote{Size ratio upper limit estimated from coverage comparison.} than 
reported in the literature for similar redshift range \citep{2013A&A...556A..55I,2013ApJ...777...18M}.

\begin{figure}
\includegraphics[width=0.45\textwidth]{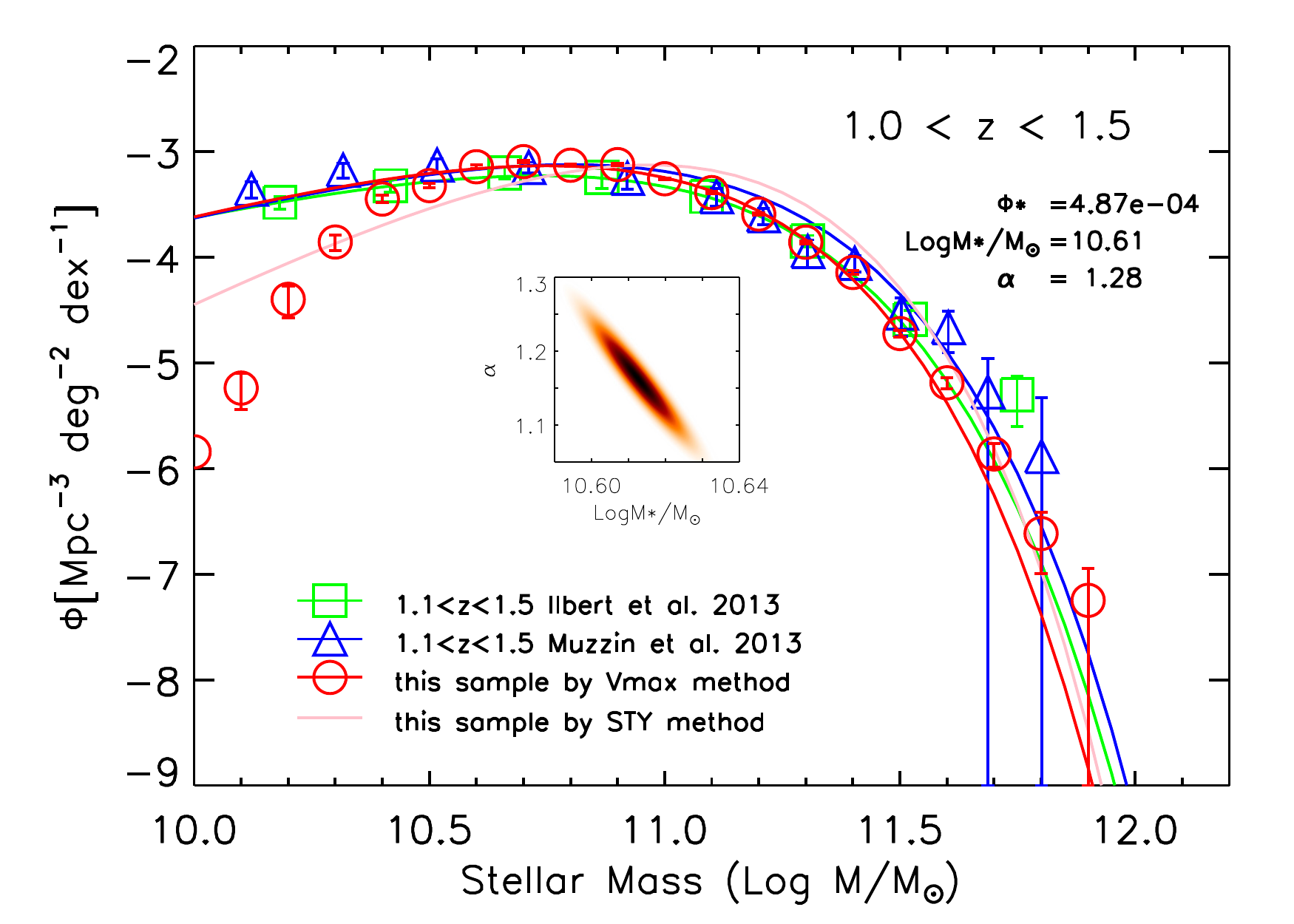}
\caption{The stellar mass function of our sample (red). 
The Schechter function fitting to our sample from Vmax method gives a
$\Phi^{\star}$=8.47e-04$\pm$1.58e-05, logM$^{\star}$=10.65$\pm$0.01, and $\alpha$=0.24$\pm$0.05,
while the STY method gives 
$\Phi^{\star}$=4.87e-04$\pm$8.29e-02, logM$^{\star}$=10.61$\pm$0.09, and $\alpha$=1.28$\pm$0.05.
The $\alpha$ vs logM$^{\star}$ probability contour is shown in the inset, 
where deeper color represents higher probability.
Green open squares are the mass function from \citet{2013A&A...556A..55I}, and 
blue open triangles are from \citet{2013ApJ...777...18M}, both in the similar redshift range as 1.1$<z<$1.5. 
Their Schechter functions are plotted using lines with same colors.
Our mass function has a drop at log(M$_*/\msun)<10.5$ 
but are much more robust at the massive end. }
\label{fig:mf}
\end{figure}

We then estimate the mass function using the V$_{max}$ method for this sample in $1.0 < z < 1.5$. 
Our mass function is plot against those obtained in COSMOS and COSMOS / UltraVISTA surveys in Figure~\ref{fig:mf}. 
Our mass function extends to the very massive end up to log(M$_*$/M$_{\odot})=$11.9.
Compared to the mass function in \citet{2013A&A...556A..55I} and \citet{2013ApJ...777...18M},
our sample have a deficit at log(M$_*/\msun)<10.5$.  
This is mainly due to the $Y<22.5$ selection in our sample 
to reject higher $z$ objects,
which was not applied in the previous studies. 
We fit a Schechter function to our sample for galaxies with log(M$_*/\msun)>10.5$, 
which yielded a $\Phi^{\star}$=8.47e-04$\pm$1.58e-05, logM$^{\star}$=10.65$\pm$0.01, and $\alpha$=0.24$\pm$0.05.
We also adopt the STY technique \citep{1979ApJ...232..352S} to 
fit a single Schechter function, 
which yields $\Phi^{\star}$=4.87e-04$\pm$8.29e-02, logM$^{\star}$=10.61$\pm$0.09, and $\alpha$=1.28$\pm$0.05.
The difference in $\Phi^{\star}$ from these two methods are 
caused by their significantly different $\alpha$ fit at the low-mass end.
The advantage of our sample is the robustness in the massive end. 
Even in the subsample with stellar mass estimates,
we have at least 10,000 galaxies at log(M$_*$/$\msun)>11$, 
and 100 galaxies with log(M$_*$/$\msun)>11.7$. 
These massive galaxies trace the very dense regions in the large scale structure, 
and our sample size permits the search for clusters and groups in their vicinities.

\section{The Large-Scale Structures Traced by this Sample} \label{sec:hod}

Massive galaxies locate in high density regions in large-scale structures. 
We use this massive galaxy sample to explore the distribution of dense regions in these fields.
For the following analysis, we only consider
galaxies with masses in log(M$_*/\msun)>11$. 
We further limit our analysis in the three fields with homogeneous Spitzer IRAC and MIPS coverages:
COSMOS, XMM-LSS, and ELAIS-N1. 
An angular two-point correlation function is derived for each field 
with the estimator $w_{obs}(\theta) = \frac{DD(\theta)-2DR(\theta)+RR(\theta)}{RR(\theta)}$ \citep{1993ApJ...412...64L}. 
Errors for each point are calculated using the Jackknife method.

\begin{figure}
\includegraphics[width=0.45\textwidth]{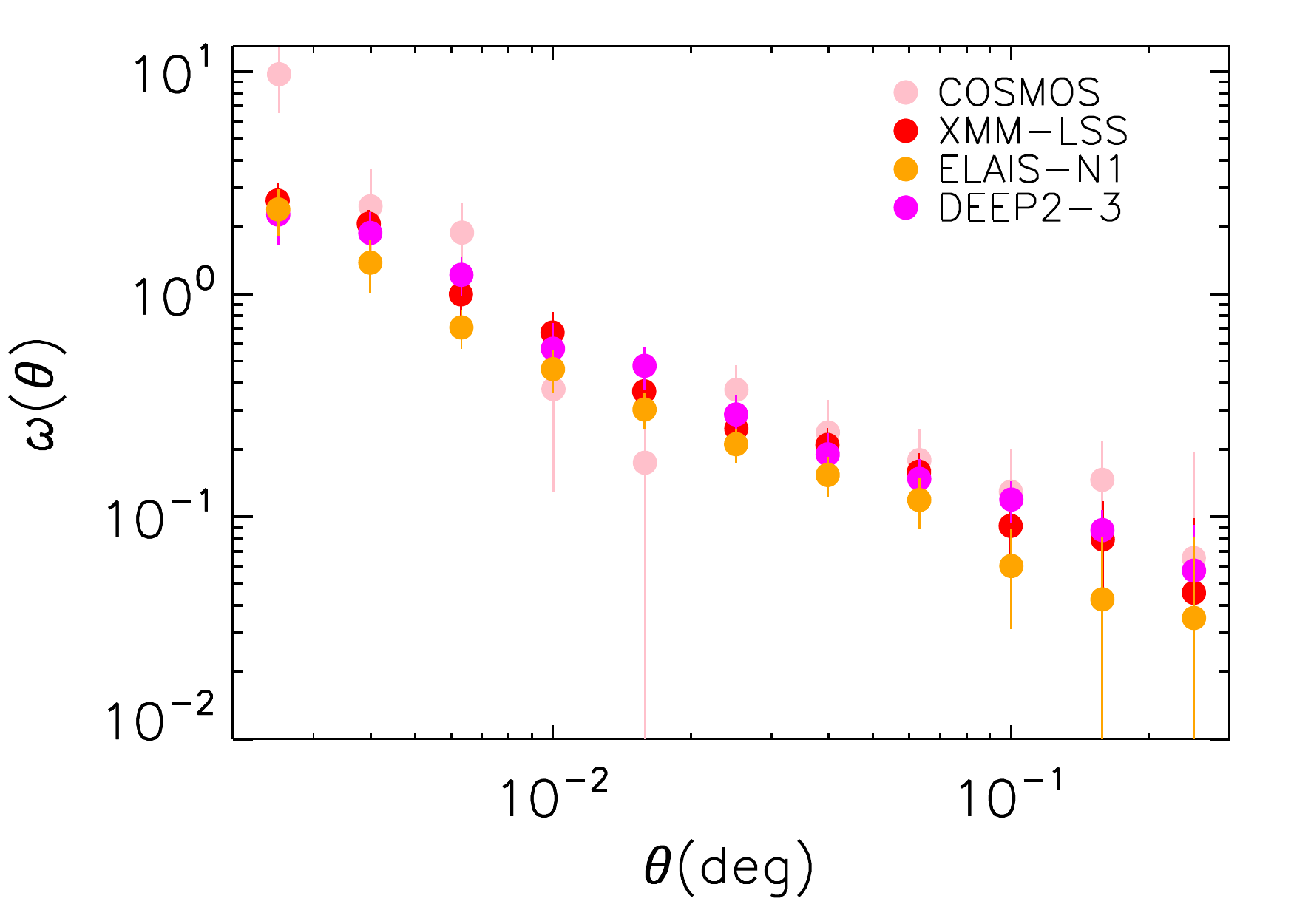}
\caption{
The two-point angular correlation functions
in the COSMOS (pink), XMM-LSS (red), ELAIS-N1(orange), 
DEEP2-3 (purple) fields.
}
\label{fig:tpcr}
\end{figure}

The angular two-point correlation functions in these three fields are plotted in Figure \ref{fig:tpcr}.
The field-to-field variation is clearly shown in this plot. 
The correlation function in the COSMOS field is the highest at small scales 
due to its high density structures at $z\sim1$ \citep{2007ApJS..172..150S}.
In larger scales, all 4 correlation functions yield 
consistent measurements within uncertainties. 

We fit a Halo Occupation Distribution (HOD) model prediction \citep{2007ApJ...667..760Z} 
to the measured correlation functions.
We constrain the parameters of our HOD model using the Markov Chain Monte Carlo (MCMC) 
method, similar to the steps described in \citet{2011ApJ...736...59Z}.
The model fitting yields a similar result for these fields: galaxy bias is 3.1$\pm$0.1 for COSMOS 
and 2.9$\pm$0.1 for the other three fields.
The mean halo mass for our sample is 
log(M$_{halo}/\msun)$$\sim$13.2.

\begin{figure}
\includegraphics[width=0.45\textwidth]{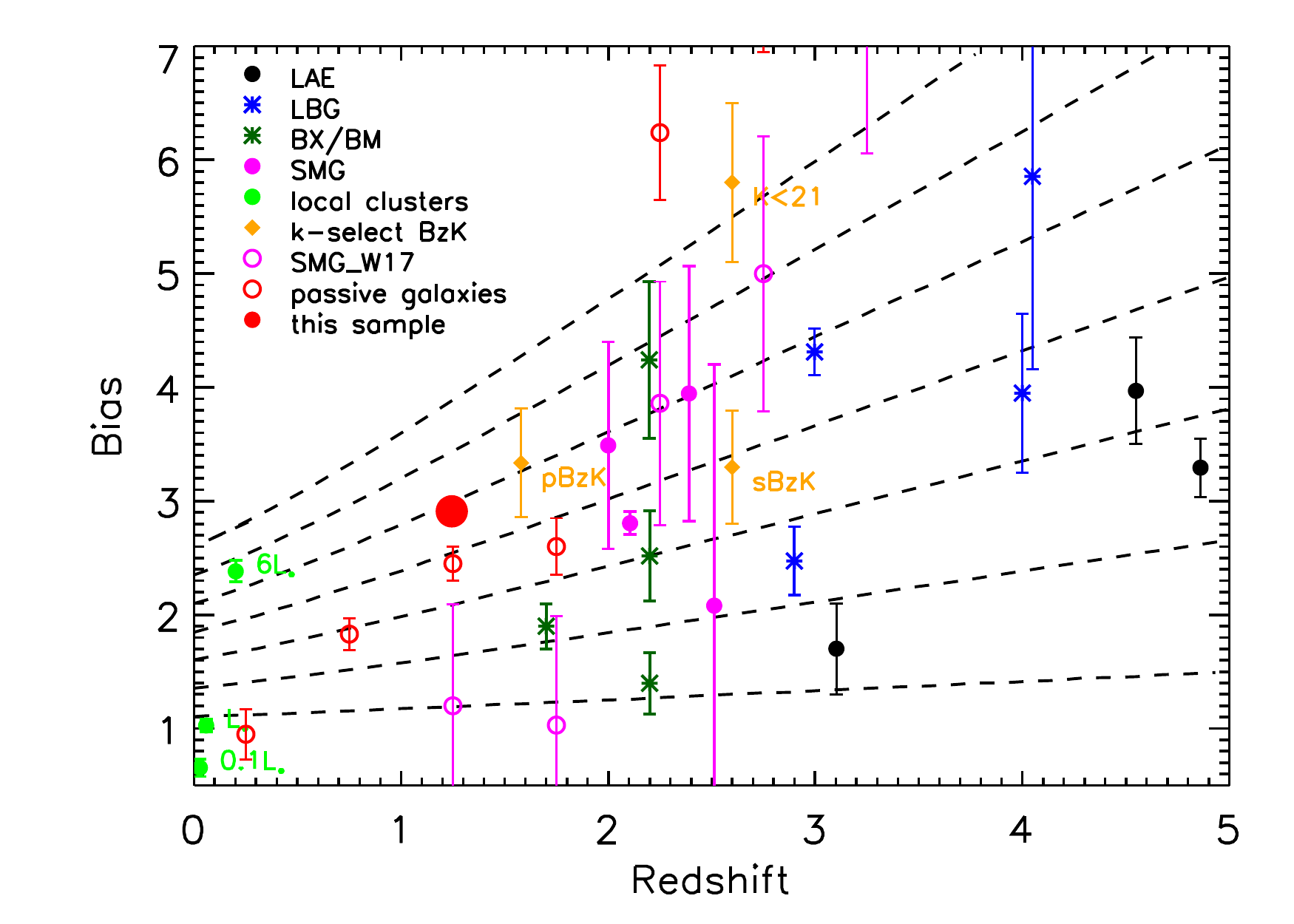}
\caption{
Bias for various galaxy populations at different redshifts. 
The red solid circles shows the mean bias of our massive passive sample (only M$>10^{11}\msun$ are plotted). 
Black solid circles are Ly$\alpha$ emission (LAE) galaxies from 
\citet{2003ApJ...582...60O, 2007ApJ...668...15K, 2007ApJ...671..278G}. 
Navy asterisks are Lyman break galaxies (LBGs) from \citep{2004ApJ...611..685O,2005ApJ...620L..75A, 2006ApJ...642...63L}.
BX/BM galaxies are the dark green asterisks \citep{2005ApJ...619..697A}.
SMGs are magenta solid circles \citep{2009ApJ...707.1201W, 2012MNRAS.421..284H, 2014ApJ...795....5O, 2015ApJ...802...64B}.
Different kinds of $Bzk$ selected galaxies are in orange solid diamonds and labelled \citep{2007ApJ...654..138Q,2008ApJ...681.1099B}.
Local galaxies from \citet{2005ApJ...630....1Z}, divided by their luminosities, are in green solid circles and 
labelled with their $L_*$ values .
Open circles are the passive galaxies (red) and SMGs (magenta) from \citet{2017MNRAS.464.1380W}.
The dash lines are the evolutionary models from \citet{1996ApJ...461L..65F},
for different halo masses, with M$_{\rm halo}$ from $10^{9}\msun$ to $10^{15}\msun$ (bottom to top).
}  
\label{fig:bias}
\end{figure}

We then compare the bias of our sample with other galaxy populations at various redshifts, 
in order to look for possible evolutionary link between populations at consecutive redshifts.
Galaxy bias at high redshifts in previous studies usually had a large uncertainty due to their sample sizes,
as shown in Figure~\ref{fig:bias}. 
The bias for our population of massive, passive galaxies (red)
has a significantly smaller uncertainty than in previous surveys at various redshifts, 
allowing a meaningful constraint on the evolutionary path.
We also use a bias evolution model, the `no-merging' model from \citet{1996ApJ...461L..65F}, 
to help understand the evolution of this population (dashed lines in Figure \ref{fig:bias}). 
Based on the model prediction, even without merging, 
our sample will evolve into a galaxy of 6$L_*$ at z$\sim$0. 
Local galaxies of such luminosity is so big that it has to be a central galaxy in a cluster. 
If looking back in time,
the model also indicates that our passive galaxy population 
may be evolutionally connected to pBzk galaxies at z$\sim$1.6 \citep{2008ApJ...681.1099B}, 
and submiliimeter galaxies (SMGs) at z$\sim$2 \citep{2009ApJ...707.1201W, 2012MNRAS.421..284H, 2014ApJ...795....5O, 2015ApJ...802...64B}. 
The evolutionary path of passive galaxies is subject 
to further studies due to the large bias uncertainties for both the pBzk and SMG populations. 
\citet{2017MNRAS.464.1380W} reported a similar conclusion by study SMG clustering properties and compared with subsequent passive galaxy populations.

\section{Summary} \label{sec:summary}
We adopt a simple color selection ($i-Y>1.3$ and $Y<22.5$) and construct
a large massive passive galaxy sample at $1.0 < z < 1.5$ in the Subaru HSC deep survey fields. 
There are a total of 33,983 galaxies in this passive galaxy sample. 
We then estimate the photometric redshifts and stellar masses for 21,132
galaxies in areas with IRAC coverages to validate our selection. 
 
Our photometric redshift estimates confirm that 
the majority ($\sim$97\%) of our sample is in the targeted redshift range ($1.0<z<1.5$), 
with a mean $z$ of 1.2, and a mean stellar mass of log(M$_*$/$\msun$)=11.
Using the subsample with photometric redshifts, we classify galaxies in the rest-frame $UVJ$ diagram 
and found $\sim$95\% are in the passive galaxy region. 
We conclude that our simple color selection with $i-Y>1.3$ and $Y<22.5$ 
yields a sample including more than 95\% of passive galaxies. 

Our morphological study on a small subsample (116 galaxies) in the CANDELS field 
also shows that the majority of our sample have an early-type morphology 
both by visual identification and in the Gini-C and Gini-M20 diagrams. 
The S\'{e}rsic index distribution shows a similar morphological mix: 
only $\sim$6\% are pure disk galaxies with n$<$1.5,
and most of them (91\%) are at 2$<$n$<$5,
implying an elliptical- or bulge$+$disk dominant profile.

Due to the bright limiting magnitude to avoid high-$z$ contaminations,
the stellar masses for the majority (93\%) of our sample are in the range of 
10.5 $<$ log(M$_*$/$\msun$) $<$ 11.5 
with a mean of log(M$_*$/$\msun$) = 11. 
The mass function derived for our sample at $1.0 < z < 1.5$ 
is consistent with those from previous studies for much smaller samples in the similar redshift range. 
Yet at the same redshift range, 
our sample is a factor of 10$\sim$17 larger at the massive end 
than existing passive galaxy samples,
resulting in a significantly more robust massive end of the mass function.
 
We use this massive sample to trace the high density regions in the large-scale structure. 
The two point angular correlation functions for the massive galaxies (log(M$_*/\msun) > $ 11)
in our sample yields a very high bias of 2.9--3.0. 
Based on the galaxy evolution tracks, 
this implies that these passive galaxies will become central galaxies in clusters at z$\sim$0, 
and are likely descendent of pBzk galaxies at z$\sim$1.6 and SMGs at z$\sim$2.

\acknowledgments

\textbf{Acknowledgments.} 
This research uses data obtained through the Telescope Access Program (TAP), 
which has been funded by the National Astronomical Observatories, Chinese Academy of Sciences.
This work is mainly based on the 
CLAUDS U band data obtained with MegaPrime/MegaCam, a joint project of CFHT and CEA/DAPNIA.
This work is also based in part on data collected at the Subaru Telescope 
and retrieved from the HSC data archive system, 
which is operated by the Subaru Telescope and Astronomy Data Center at National Astronomical Observatory of Japan.
This work is based in part on observations made
with the $Spitzer$ Space Telescope, which is operated by the Jet Propulsion Laboratory,
California Institute of Technology under a contract with NASA. 

The authors would like to thank Kevin Xu,  Haojing Yan, Chuan He, and Zijian Li for their helpful discussions.
The authors would also like to thank the anonymous referee who helped to improve the 
content and presentation of this paper. 
Support for this work is provided by the Chinese National
Nature Science foundation (NSFC) grant number 10878003, and also
supported in part by the National Key R\&D Program of China via
grant number 2017YFA0402703 and by NSFC grants 11433003, 11933003,
11373034, 11803044 and 11673028. 
Y.-S.~Dai acknowledges the support from
Chinese Academy of Sciences for their support under the `Bairen' Funding program. 
Additional support came from the Chinese Academy of
Sciences (CAS) through a grant to the South America Center for
Astronomy (CASSACA) in Santiago, Chile.

\end{CJK*}
\end{document}